\documentclass[12pt]{iopart}

\usepackage{iopams}
\usepackage{graphicx}
\usepackage{amssymb}
\usepackage{cite}
\usepackage{dcolumn}
\usepackage{bm}
\usepackage{footnote}
\usepackage{wasysym}
\usepackage{color}
\usepackage{multirow}
\usepackage{bm}
\usepackage{sidecap}
\usepackage{ulem}
\usepackage{url}
\usepackage{textcase}

\begin{document}

\title[Modified empirical formulas and machine learning for $\alpha$-decay systematics]{Modified empirical formulas and machine learning for $\alpha$-decay systematics}

\author{G. Saxena$^{1,\dagger}$, P. K. Sharma$^{2}$ and Prafulla Saxena$^{3}$}

\address{$^{1}$Department of Physics (H\&S), Govt. Women Engineering College, Ajmer-305002, India}
\address{$^{2}$Govt. Polytechnic College, Rajsamand-313324, India}
\address{$^{3}$Department of Computer Science and Engineering, Indian Institute of Technology, Kanpur-208016, India.}
\ead{$^{\dagger}$gauravphy@gmail.com}
\vspace{10pt}
\begin{indented}
\item[]September 2020
\end{indented}

\begin{abstract}
Latest experimental and evaluated $\alpha$-decay half-lives between 82$\leq$Z$\leq$118 have been used to modify two empirical formulas: (i) Horoi scaling law [J. Phys. G \textbf{30}, 945 (2004)], and Sobiczewski formula [Acta Phys. Pol. B \textbf{36}, 3095 (2005)] by adding asymmetry dependent terms ($I$ and $I^2$) and refitting of the coefficients. The results of these modified formulas are found with significant improvement while compared with other 21 formulas, and, therefore, are used to predict $\alpha$-decay half-lives with more precision in the unknown superheavy region. The formula of spontaneous fission (SF) half-life proposed by Bao \textit{et al.} [J. Phys. G \textbf{42}, 085101 (2015)] is further modified by using ground-state shell-plus-pairing correction taken from FRDM-2012 and using latest experimental and evaluated spontaneous fission half-lives between 82$\leq$Z$\leq$118. Using these modified formulas, contest between $\alpha$-decay and SF is probed for the nuclei within the range  112$\leq$Z$\leq$118 and consequently probable half-lives and decay modes are estimated. Potential decay chains of $^{286-302}$Og and $^{287-303}$119 (168$\leq$N$\leq$184: island of stability) are analyzed which are found in excellent agreement with available experimental data. In addition, four different machine learning models: XGBoost, Random Forest (RF), Decision Trees (DTs), and Multilayer Perceptron (MLP) neural network are used to train a predictor for $\alpha$-decay and SF half-lives prediction. The prediction of decay modes using XGBoost and MLP are found in excellent agreement with available experimental decay modes along with our predictions obtained by above mentioned modified formulas.
\end{abstract}

\submitto{\JPG}

\section{Introduction}

The journey of $\alpha$-decay, starting from Rutherford followed by Geiger-Nuttall law \cite{geiger1911} in 1911 along with quantum mechanical formulation by Gamow \cite{gamow1928} and, Guerney and Condon \cite{condon1928} in 1928 to till this year 2020 \cite{cai2020,deng2020,adamian2020,ghodsi2020,voinov2020} dominates the investigations of superheavy nuclei in such a way that most of the experimental and theoretical studies of superheavy nuclei are concentric primarily on the $\alpha$-decay mode. Study of $\alpha$-decay chains plays a crucial role for synthesis of new elements in the laboratory and also to investigate the nuclear structure properties \cite{hofmann2000,hamilton2013,ogan2015,heenen2015,oganrpp2015,oganpt2015,hofmann2016,dull2018,nazar2018,giuliani2019}. The synthesis of superheavy elements in laboratories at GSI \cite{hofmann2000,Hofmann2011}, RIKEN \cite{Morita2007} and JINR \cite{hamilton2013,Oganessian2010,Oganessian2015} is mainly governed by cold-fusion reactions between $^{208}$Pb, $^{209}$Bi and beams of A$>$50 \cite{hamilton2013} or hot-fusion reactions between long-lived actinide nuclei from $^{238}$U to $^{249}$Cf and $^{48}$Ca beams \cite{ogan2015}. As an extension of the periodic chart, decay chains of $^{293}$117
and $^{294}$117 have been reported by fusion reactions between $^{48}$Ca and $^{249}$Bk with 11 new nuclei \cite{Oganessian2010} and also of $^{294}$118 by $^{249}$Cf+$^{48}$Ca reaction \cite{ogan2006}. In addition, afterward, eleven new heaviest isotopes of elements Z$=$105 to Z$=$117 are identified validating the concept of the island of enhanced stability for superheavy nuclei (SHN) \cite{ogan2011}. For the synthesis of elements with Z$\geq$118 the attempts and progress are still going on and in near future few more SHN are expected to be detected \cite{hofmann2016,Oganessian2009}. \par

In recent times, experimental works are performed on the Dubna gas-filled magnetic recoil separator (GFS) to study the reactions of the formation of $^{294}$Ts and $^{294}$Og nuclides along with future possibilities of synthesizing new elements with Z$=$119 and Z$=$120 \cite{voinov2020}. Similarly, the observation of a new decay chain of $^{294}$Og along with prospects for reaching new isotopes $^{295,296}$Og are reported \cite{brewer2018}. Many properties of the decay of the
heaviest nuclei with Z$=$112–118 indicate a considerable stabilizing effect upon approaching the island of stability N$\sim$184 \cite{ogan2015}. However, up to now the heaviest experimentally known nuclei $^{294}$Ts and $^{294}$Og have neutron number N$=$177 and N$=$176, respectively. Therefore, future experiments are expected to synthesize heavier elements which may have the neutron number between N$\sim$170 and N$\sim$184. With this in view, the study of decay chains of Z$=$118 and 119 with neutron number 168$\leq$N$\leq$184 may be of extreme importance and crucial for experimental point of view, which is precisely the aim of this article. \par

The study of $\alpha$-decay chains consists of the determination of $\alpha$-decay half-life ($log_{10}T_{1/2}$) which is first formulated as linear dependence on reciprocal of the square root of $\alpha$-decay energy ($\sqrt{Q}$) \cite{geiger1911}. In 1966, Viola and Seaborg \cite{vss1966} predicted a simple formula that is based on the Gamow model by including the intercept parameters of linear dependence on the charge number of the daughter nucleus and also modified in Refs. \cite{sobi1989,sobiczewski2005}. Afterwards, Brown \cite{brown1992}, Royer \cite{royer2000} and Ren \textit{et al.} \cite{renA2004} have formulated the empirical formulas by studying the experimental variation of the logarithmic half-lives. Later on, Ni \textit{et al.} \cite{nrdx2008} have derived the empirical formula (NRDX formula) for alpha decay half-lives from the WKB barrier penetration probability with certain approximations. Likewise, Qi \textit{et al.} \cite{qi2009} proposed a linear universal decay law (UDL) from the microscopic mechanism of the charged-particle emission. In addition to the formulas mentioned above, some semi-empirical relationships including microscopic effects, such as the universal (UNIV) curve, and a semi-empirical formula based on fission theory (SemFIS), were proposed by Poenaru \textit{et al.} which are subsequently modified in Refs. \cite{poenaru2012} and \cite{dong2010} proposing new names NUNIV and SemFIS2, respectively. Recently, many of these formulas have been modified by adding asymmetry terms and referred as modified Royer formula (Akre formula) \cite{akrawy2017}, modified UDL formula (MUDL) \cite{akrawy2019}, modified scaling law of Brown (MSLB) \cite{akrawyprc2019}, modified Viola-Seaborg formula (MVS) \cite{akrawyprc2019}, and new Ren A formula \cite{newrenA2019}. Few of the above mentioned formulas are refitted using latest experimental alpha decay half-lives viz. (i) modified Royer GLDM (Rm), modified Viola-Seaborg (VSm1 and VSm2) and modified Sobiczewski-Parkhomenko (SPm1 and SPm2) formulas \cite{dasgupta2009}, (ii) modified version of the Brown empirical formula (mB1 and mB2) \cite{budaca2016} (iii) new universal decay law (NUDL), new Royer (NRo) and new Sobiczewski-Parkhomenko (NSP) formulas \cite{zhao2018}. Moreover, Manjunatha \textit{et al.} have presented a formula \cite{manjunatha2019} based on second-order polynomial fitting of $Z^{0.4}_{d}/\sqrt{Q}$ which is subsequently modified in one of our recent work \cite{singh2020}. \par

Besides the above-mentioned formulas, in 2004, Horoi \textit{et al.} \cite{horoi2004} have proposed the scaling law for cluster decay and in 2005, Parkhomenko and Sobiczewski \cite{sobiczewski2005} have included excitation energy for odd-A and odd-odd nuclei in an empirical formula for Z$>$82. Both these formulas may be modified further and the parameters may be adjusted with the fitting of updated experimental data which is one of the motives of the present work.\par

Another decay mode that competes with $\alpha$-decay and acts as a decider to terminate the $\alpha$-decay chains is spontaneous fission (SF). The first semi-empirical formula for SF was proposed by Swiatecki \cite{swiatecki} in 1955 and afterward in 2005, Ren \textit{et al.} \cite{ren2005} have generalized formula for spontaneous fission half-lives of even-even nuclei in their ground state to both the case of odd nuclei and the case of fission isomers. Later on,
Xu \textit{et al.} \cite{xu2008} have proposed empirical formula of SF for even-even nuclei which is followed by the semi-empirical formula given by Santosh \textit{et al.} \cite{santosh2010}. In 2012, Karpov \textit{et al.} \cite{karpov2012} have proposed a new formula based on barrier height which is modified with even-odd corrections by Silisteanu \textit{et al.} \cite{silisteanu2015}. In 2015, a modified formula is proposed by Bao \textit{et al.} \cite{bao2015} for determining the spontaneous fission half-lives based on Swiatecki’s formula, including the microscopic shell correction from FRDM-1995 \cite{moller1995} and isospin effect. This formula was able to reproduce the experimental data quite well, however, it may be further modified with shell plus pairing correction from FRDM-2012 \cite{moller2012} which is also an object of present work.\par

In recent times, machine learning (ML) algorithms \cite{lecun2015,schmidhuber2015} have participated as an alternate and powerful tool to study and predict a complex data of physics \cite{carleo2019,mehta2019,guest2018}. In particular, machine learning methods are employed in nuclear physics for prediction of nuclear masses \cite{athan2004,niu2018}, binding energies \cite{tuhin2020}, $\beta$-decay \cite{costiris2009} and charge radii \cite{utama2016}. Since the predictions in the superheavy realm by various theories and empirical/semi-empirical formulas are still not so precise, therefore, machine learning methods could be very useful to predict decay modes, decay chains, and half-lives of unknown territory of superheavy nuclei. This possibility has invoked us to apply machine learning methods for the prediction of $\alpha$-decay, SF and for the determination of half-lives of known and unknown nuclei to visualize the merit of this alternate approach in superheavy region.\par

The paper is organized as follows. In section II we provide the original formulas and their modified versions with concerned coefficients for $\alpha$-decay and spontaneous fission (SF) half-lives. These formulas are fitted using $scipy.optimize.curve\_fit$ library \cite{scipy} along with the Levenberg-Marquardt algorithm which is used for non-linear least-squares fit \cite{wood}. In Section III we give a brief outline of the all machine learning algorithms employed. In section IV we first compare results of modified formulas with already established formulas and then compare our results of half-lives for a few experimentally known decay chains. Then, we predict probable decay modes for the nuclei in the range 112$\leq$Z$\leq$118 which is followed by a prediction of potential decay chains of $^{286-302}$Og and $^{287-303}$119.\par

\section{Formalism for $\alpha$-decay and spontaneous fission (SF) half-lives}

\subsection{Horoi Formula for $\alpha$-decay}
In 2004, Horoi \textit{et al.} \cite{horoi2004} have generalized scaling law for the decay time of alpha particles for cluster decay and proposed that the $log_{10}$T$_{1/2}$ depends linearly on the scaling variable S = (Z$_{c}$Z$_{d}$)$^{0.6}$/$\sqrt{Q}_c$ and on the square root of the reduced mass ($\mu$) of cluster and daughter. From a detailed analysis, he proposed a model-independent law for the entire set of cluster decay data of the following form:
\begin{eqnarray}\label{horoi}
log_{10}T_{1/2} &=& (a_1 \sqrt{\mu} + b_1)[(Z_c Z_d )^{0.6} Q^{-1/2} - 7] + (a_2 \sqrt{\mu} + b_2)
\end{eqnarray}
here, Z$_c$ and Z$_d$ are the charges of the cluster and daughter nuclei, $\mu$ is the reduced mass of the cluster–daughter system.
\subsection{Modified Horoi Formula (MHF-2020) for $\alpha$-decay}
In the present work, we modify the Horoi formula \cite{horoi2004} for $\alpha$-decay by adding asymmetry dependent terms ($I$ and $I^2$) which are linearly related to the logarithm of $\alpha$-decay half-lives \cite{akrawy2019}. This kind of modification has been successfully established by few other studies \cite{akrawyprc2019,akrawynpa2018,newrenA2019} and one of our recent study \cite{singh2020} in which the addition of $I$ and $I^2$ is found in a good match with experimental systematics. The modified Horoi formula (MHF-2020) is given by:
\begin{eqnarray}\label{horoi-modi}
log_{10}T_{1/2} &=& (a \sqrt{\mu} + b)[(Z_c Z_d )^{0.6} Q^{-1/2} - 7] + (c \sqrt{\mu} + d) + eI + fI^2
\end{eqnarray}

Here, new coefficients are obtained by fitting the half-lives from the latest evaluated database \cite{nndc} of 366 nuclei in the range 82$\leq$Z$\leq$118. The set of coefficients which is obtained by fitting 103 even-even (e-e), 101 even-odd (e-o), 76 odd-even (o-e) and 86 odd-odd (o-o) data separately is the set 1 for which coefficients are given in the Table \ref{coeff-horoi-set1} along with the sample size in each category. Whereas, set 2 is obtained by fitting of all the data together and the coefficients are given in the Table \ref{coeff-horoi-set2}. Set 1 comprises total 24 coefficients and can be used separately for all even(odd)-odd(even) combinations of Z-N, whereas set 2 contains only 9 coefficients, out of which 5 coefficients are same for all sets of nuclei and only 1 coefficient is different for even(odd)-odd(even) combinations of Z-N (refer to Tables \ref{coeff-horoi-set1} \& \ref{coeff-horoi-set2}).

\begin{table}[!htbp]
\caption{The coefficients of modified Horoi formula (MHF-2020: Set 1).}
\centering
\resizebox{1.0\textwidth}{!}{%
\begin{tabular}{l@{\hskip 0.6in}c@{\hskip 0.6in}c@{\hskip 0.6in}c@{\hskip 0.6in}c@{\hskip 0.6in}c@{\hskip 0.6in}c}
\hline
\hline
Set 1       &    a      &         b      &     c   &         d&      e&        f\\
\hline
Even-even(103)  &  834.367& -1648.187	&      -875.905   &    1722.168   &    90.116     &     -267.214   \\
Even-odd(101)   &  425.941& -837.910  &      -349.093   &    683.399     &    23.602     &     -79.824     \\
Odd-even(76)   &  -32.607& 71.208	   &      206.158    &    -411.086    &    -47.871    &     116.354     \\
Odd-odd(86)    &  145.412&-282.793	  &      45.602     &    -90.823  &    -53.034    &     128.653    \\
\hline
\hline
\end{tabular}}
\label{coeff-horoi-set1}
\end{table}

\begin{table}[!htbp]
\caption{The coefficients of modified Horoi formula (MHF-2020: Set 2).}
\centering
\resizebox{0.9\textwidth}{!}{%
\begin{tabular}{c@{\hskip 0.6in}c@{\hskip 0.6in}c@{\hskip 0.6in}c@{\hskip 0.6in}c@{\hskip 0.6in}c}
\hline
\hline
Set 2       &    a      &         b      &     c   &       e&        f\\
\hline
For all nuclei&643.540& -1269.906 &-480.259&33.377&-116.008\\
\cline{2-6}
\cline{2-6}
&&Even-even&Even-odd&Odd-even&Odd-odd\\
\cline{3-6}
&d&943.118&943.534&943.440&943.810\\
\hline
\end{tabular}}
\label{coeff-horoi-set2}
\end{table}

\subsection{Sobiczewski formula for $\alpha$-decay}
In 2005, Parkhomenko and Sobiczewski \cite{sobiczewski2005} proposed a phenomenological formula for the description of $\alpha$-decay half-lives for the nuclei heavier than $^{208}$Pb which is referred as Sobiczewski formula and given by:

\begin{equation}\label{sobic}
log_{10}T_{1/2} = aZ(Q_{\alpha} - \overline{E}_i)^{-1/2} + bZ + c
\end{equation}


\subsection{Modified Sobiczewski formula (MSF-2020) for $\alpha$-decay}
In the present work, we modify the Sobiczewski formula \cite{sobiczewski2005} for $\alpha$-decay by adding asymmetry dependent terms ($I$ and $I^2$) which are linearly related to the logarithm of $\alpha$-decay half-lives \cite{akrawy2019}. The modified Sobiczewski formula (MSF-2020) is given by:

\begin{equation}\label{sobic-modi}
log_{10}T_{1/2} = aZ(Q_{\alpha} - \overline{E}_i)^{-1/2} + bZ + c + dI +eI^2
\end{equation}

As mentioned above, the new coefficients are obtained by fitting the half-lives from the latest evaluated database \cite{nndc} of 366 nuclei in the range 82$\leq$Z$\leq$118 and for this modified formula, coefficients are given in the Table \ref{coeff-sobic}.

\begin{table}[!htbp]
\caption{The coefficients of modified Sobiczewski formula (MSF-2020).}
\centering
\resizebox{0.85\textwidth}{!}{%
\begin{tabular}{l@{\hskip 0.5in}c@{\hskip 0.5in}c@{\hskip 0.5in}c@{\hskip 0.5in}c@{\hskip 0.5in}c@{\hskip 0.5in}c}
\hline
\hline
Set        & E$_i$    &  a      &         b    &     c     &         d     &      e\\
\hline
Even-even  & 0.000 &1.266 &  -0.144   &  -23.304 & -81.216    &  275.650  \\
Even-odd   & 0.171 &1.539 &  -0.188   &  -32.353 & -19.488    &  56.520  \\
Odd-even   & 0.113 &1.566 &  -0.171   &  -32.748 & -38.831    &  99.568  \\
Odd-odd    & 0.284 &1.265 &  -0.130   &  -25.116 & -58.267    &  157.311  \\
\hline
\hline
\end{tabular}}
\label{coeff-sobic}
\end{table}

\subsection{Bao formula for SF}
In 2015, Bao \textit{et al.} \cite{bao2015} have modified Swiatecki’s formula \cite{swiatecki} of spontaneous fission half-lives considering role of shell effects.
Taking into account the dependence of half-lives on the shell correction and isospin effect, the formula was given as:

\begin{eqnarray}\label{baoSF}
log_{10}T_{1/2}^{SF} (yr)= c_1 + c_2 \left(\frac{Z^2}{(1-kI^2)A}\right) + c_3 \left(\frac{Z^2}{(1-kI^2)A}\right)^2 + c_4 E_{sh} + h_i
\end{eqnarray}

Values of the parameters are: c$_1$$=$1174.353441, c$_2$$=$$-$47.666855, c$_3$$=$0.471307, c$_4$$=$3.378848 and k$=$2.6. The values of h$_i$ for various sets of nuclei are: h$_{e-e}$$=$0, h$_{e-o}$$=$2.609374, h$_{o-e}$$=$2.619768, h$_{o-o}$$=$h$_{e-o}$+h$_{o-e}$$=$5.229142. E$_{sh}$ is the shell correction energy taken from FRDM-1995 \cite{moller1995}.

\subsection{Modified Bao formula (MBF-2020) for SF}
For the first time, recently, it has been shown by Poenaru \textit{et al.} \cite{poenaru2017,poenaru2017epl} that in a spontaneous fission process the shell plus pairing corrections calculated with Strutinsky’s procedure may give a strong argument for preformation of a light fission fragment near the nuclear surface. With this in view, in the present work, we have modified the Bao formula \cite{bao2015} by introducing shell plus pairing correction energy taken from FRDM-2012 \cite{moller2012} and fitting with latest evaluated spontaneous fission half-lives \cite{nndc}. The modified Bao formula (MBF-2020) is given by:

\begin{eqnarray}\label{baoSF}
log_{10}T_{1/2}^{SF} (s) &=& c_1 + c_2 \left(\frac{Z^2}{(1-kI^2)A}\right) + c_3 \left(\frac{Z^2}{(1-kI^2)A}\right)^2 + c_4 E_{s+p}
\end{eqnarray}

Here k$=$2.6 and other coefficients are readjusted and fitted by considering 49 nuclei in the range 82$\leq$Z$\leq$118 and are c$_2$$=$$-$37.051, c$_3$$=$0.374, c$_4$$=$3.110. The values of c$_1$ for various sets of nuclei are c$_1$(e-e)$=$893.264, c$_1$(e-o)$=$895.415, c$_1$(o-e)$=$896.845 and c$_1$(o-o)$=$897.019.

\section{Machine learning methods}
\subsection{Introduction to Machine learning}
              Machine learning (ML) is a field of study which explores algorithms where the machine can learn itself from data. ML methods learn hidden patterns, rules, and make intelligent decisions. Typical ML methods leverage optimization strategies that iteratively learns a model by minimizing loss calculated on target or actual ground truth during the training phase. Machine learning methods are categorized into various sub-categories based on their learning style.
              \newline \textbf{Supervised learning:} Supervised learning methods learn a function that maps input to an output based on example input-output pairs. In other words, labels or ground-truths are required in these methods, which leads to model enhancement by minimizing loss incurred in previous iterations. For example, Random forest (RF), XGBoost and decision tree (DT) are based on supervised learning.
              \newline \textbf{Unsupervised learning:} Unsupervised learning methods follow algorithms which groups input data based on their similarity. No label is required for unsupervised learning as input features are processed to analyze the similarity between data.
              \newline \textbf{Online learning:} Data is increasing continuously, where the older trained model may no longer be good enough for newer data with different patterns as the model does not learn their properties. Online learning approaches incrementally update the model, which can improve model performance for newer samples with different patterns.
              \newline \textbf{Ensemble learning:} The ensemble method leverages multiple learning algorithms to obtain better predictive performance. Many weak learners like decision trees play their part, and a collective decision is taken to get an appropriate prediction. Random forest and XGBoost methods are ensemble learning methods we have used in this work. Bagging and boosting strategies are used by ensemble learning methods to train multiple decision models. Techniques used in ensemble learning, i.e. bagging and others, tend to reduce problems related to over-fitting. Further, in small-sized datasets, these methods are more beneficial than neural networks in terms of training time and accuracy.
              \newline \textbf{Model Training:} Machine learning methods leverages optimization methods which iteratively learn hidden data pattern by updating on previous learning model. The goal of optimization methods, as well as the final model, is to minimize training loss. The model is trained for some set of epochs until loss does not decrease further. To validate the learned model, validation and test set are kept separate from the training set. K-fold cross-validation is one of the highly practiced validation approaches. According to K-fold cross-validation, data is partitioned into K folds. Each fold is treated as a validation set, and the remaining folds are considered for training. This methodology is well suited with small-sized data as the model may or may not perform well on the specific or random split. We have used a 5-fold cross-validation approach in this work.
              ML problems or task are broadly divided into two categories known as \textbf{Classification} and \textbf{Regression}.
              In Classification problems, the learned model classifies input among a set of predefined classes, i.e., predicting a vehicle class by looking at its features.  Whereas, in Regression, task input is bounded by values. It predicts the possible value of the given input, i.e., predicting a person's age by looking at physical features.
\subsection{Machine learning algorithms}
              In this work, we have used Supervised learning methods and framed our problem as a Regression one. We use three tree-based supervised ML methods and one neural network method. Considering the requirement of predicting life time, which is a real value, the problem becomes a Regression problem.

              \subsubsection{Decision Trees}
              A decision tree (DT) \cite{DTs} uses a tree-like data structure to learn the hidden data pattern based on some set of rules. Decisions are taken at tree nodes based on attribute importance. The importance of attributes is decided by some metrics, i.e. entropy or gini coefficients. The final decision is taken at the leaf of the tree. The decision tree approach is similar to conditional rule-based decision-making, but it is highly prone to overfitting with large datasets. As a single tree may overfit on some set of samples..
              \subsubsection{Random Forest}
              Random Forest (RF) \cite{rf} algorithm is an ensemble learning technique in which the number of uncorrelated decision trees are constructed to avoid overfitting to the training data. RF uses bagging techniques in which a random subset of features (attributes) and data are sampled to fit various decision trees. It creates a forest of decision trees. A collective decision is then taken from the generated forest for a final prediction.

              \subsubsection{Multi-layer perceptrons}
              \begin{figure}[htbp]
                     \centering
                     \includegraphics[width=0.5\textwidth]{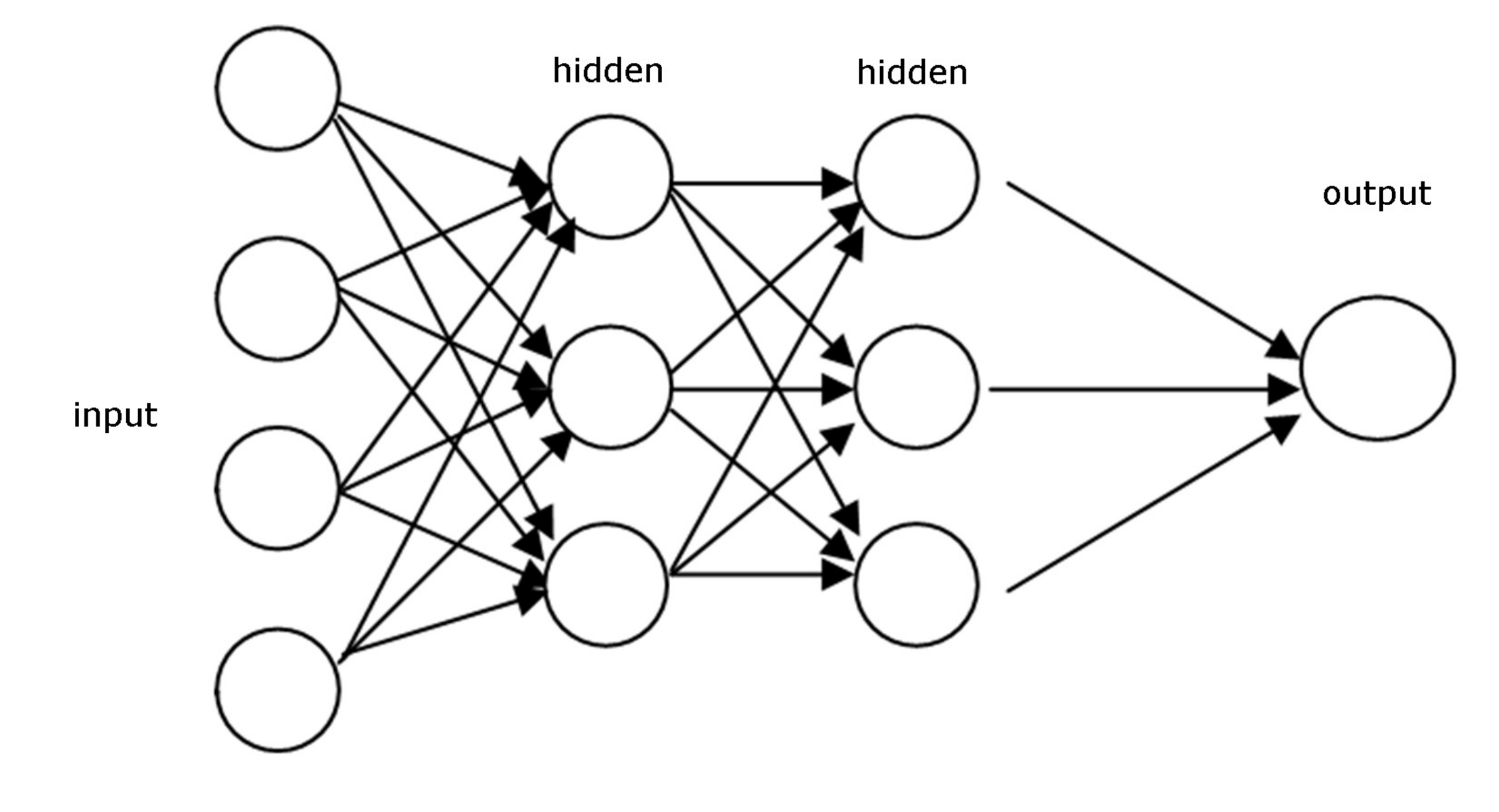}
                     \caption{Multi-layer perceptron.}\label{fig1-mlp}
              \end{figure}
              Multilayer perceptron (MLP)\cite{mlp} belongs to a class of feed-forward artificial neural network (ANN). It uses a supervised learning technique called backpropagation for training. MLP contains three layers named input layers, hidden layer(s), and output layer, as shown in Fig. \ref{fig1-mlp}. Even MLPs with a single hidden layer are powerful models that can approximate a continuous function under some constraints. Hidden layers can be increased to learn a complex function best suited to training data. An activation function is used at hidden layers to introduce non-linearity to distinguish data that is not linearly separable. In backpropagation based on error optimization, weights of hidden layers are adjusted iteratively. MLPs are powerful to approximate any kind of function based on training data but highly prone to overfitting due to complex model structure and less data availability.
              \subsubsection{XGBoost}
              XGBoost \cite{xgboost} stands for eXtreme Gradient Boosting, an ensemble machine learning algorithm that uses a gradient boosting framework. XGBoost uses gradient boosted decision trees designed for performance and speed with parallel tree boosting. It supports classification, regression, and various rank objective functions that provide a more efficient solution than its counterparts like Decision trees and Random forest. Gradient boosting is an ensemble learning technique in which many weak learners or models (i.e., Decision Trees) are combined to produce a final decision. All learned models and their weighted sum are taken into consideration for the final results. For large and unstructured datasets, artificial neural networks usually outperform all other algorithms. However, with small-to-medium structured or low-dimensional data, neural networks suffer in performance where XGBoost (decision tree-based algorithms) perform well and considered best-in-class.

\subsection{Employed approach}
 \begin{figure}[!htbp]
                     \centering
                     \includegraphics[width=0.7\textwidth]{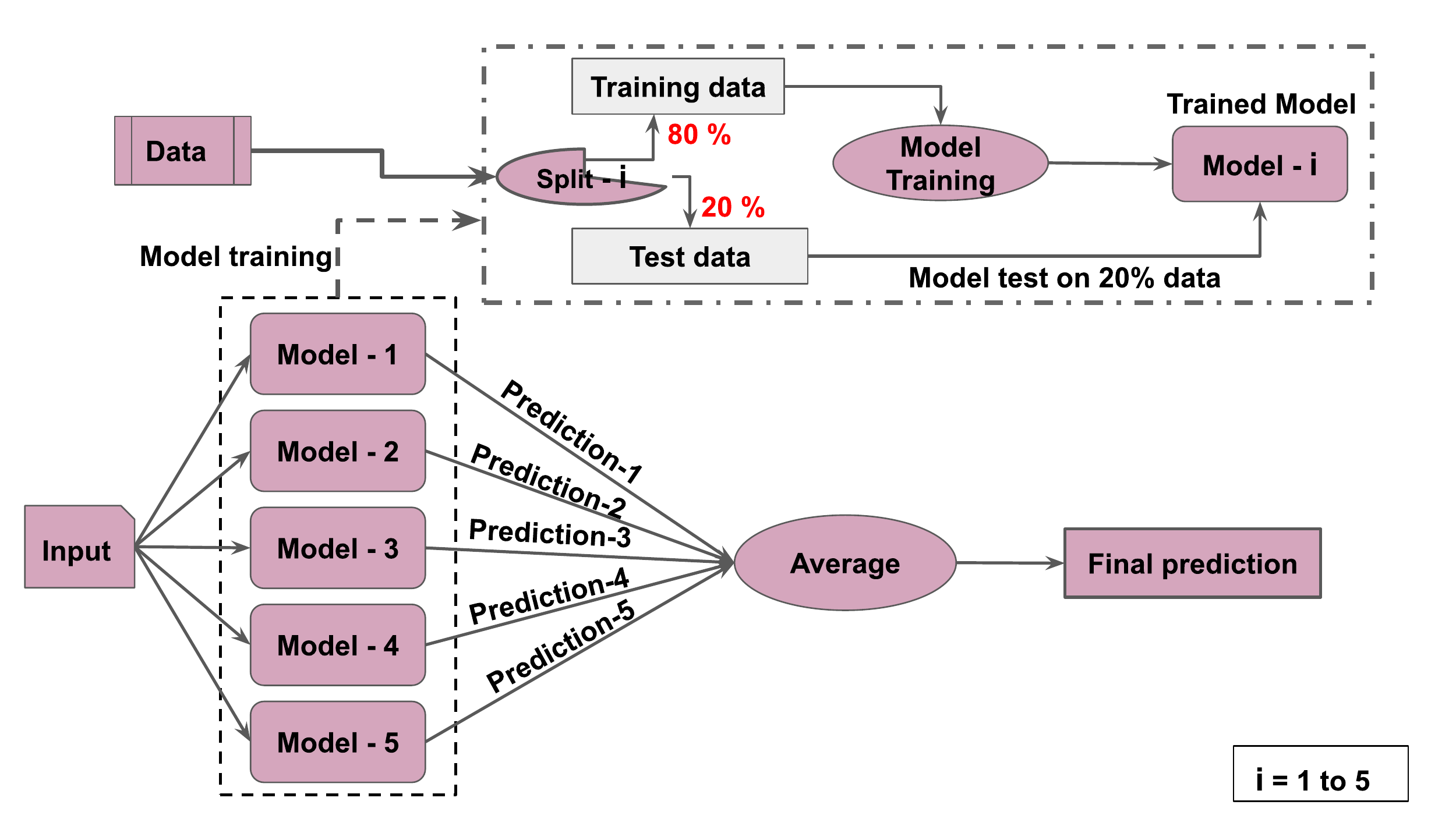}
                     \caption{Flow of model training and prediction of final result.}\label{fig2-approach}
              \end{figure}

              We have used K-fold cross-validation in all machine learning models and reported the average RMSE and mean deviation in Table \ref{ML-RMSE}. Considering 5-fold cross validation, models are trained and tested five times so that each group gets a chance to be the test set. For one group, the training set consists 80\% of the dataset, and the remaining 20\% is used for the test. We separate prediction data from training data and merge it with some newer samples for which we do not have experimental results. We use XGBoost and MLP models for alpha data prediction. For SF data prediction we use XGBoost and Random forest. We take 5 prediction from 5 models trained during 5-fold cross validation. The average of all five predictions is reported as the final answer. The reason for using five model average is that a single model may give excellent performance on a specific input data if the model properly learns its similar features, but if the model does not learn similar embedding, then it may drastically degrade the prediction result whether it is possible that other model trained on different split may give a more accurate prediction. To overcome the issue of best prediction due to biased training and worst prediction due to improper training, we use an average of all five splits. A detailed flow of our approach is shown in Fig. \ref{fig2-approach}.\\

\section{Results and discussions}
In the beginning, to test the predictive power of few known and mostly used $\alpha$-decay empirical/semi-empirical formulas, we have calculated $\alpha$-decay half-lives using experimental and evaluated Q$_{\alpha}$ \cite{nndc} for 366 nuclei within the range 82$\leq$Z$\leq$118. The formulas used are Sobiczewski \cite{sobiczewski2005}, NRDX \cite{nrdx2008}, new universal curve (NUNIV) \cite{poenaru2012},  semi-empirical formula based on fission theory (SemFIS2) \cite{dong2010}, Akre formula given by Akrawy \textit{et al.} \cite{akrawy2017}, modified UDL formula \cite{akrawy2019}, MSLB formula \cite{akrawyprc2019}, modified Viola-Seaborg formula \cite{akrawyprc2019}, Manjunatha present formula \cite{manjunatha2019}, Horoi \cite{horoi2004}, new Ren A formula \cite{newrenA2019}, modified Royer GLDM (Rm) \cite{dasgupta2009}, modified Viola-Seaborg (VSm1 and VSm2) \cite{dasgupta2009}, modified Sobiczewski-Parkhomenko (SPm1 and SPm2) formulas \cite{dasgupta2009}, modified version of the Brown empirical formula (mB1 and mB2) \cite{budaca2016}, new universal decay law (NUDL), new Royer (NRo) and new Sobiczewski-Parkhomenko (NSP) formula \cite{zhao2018}. In addition, the modified formulas in the present work i.e. modified Horoi formula (MHF-2020) (set-1 and set-2) and modified Sobiczewski formula (MSF-2020) are also tested for the same set of nuclei. \par

\begin{table}[!htbp]
\caption{Root mean square error (RMSE), mean deviation ($\overline{\delta}$), index F$=$10$^{\overline{\delta}}$ (in the unit of second) and No. of coefficients of various formulas for $\alpha$-decay half-lives calculated for nuclei in the range with 82$\leq$Z$\leq$118.}
\centering
\resizebox{1.0\textwidth}{!}{%
\begin{tabular}{l@{\hskip 0.5in}c@{\hskip 0.5in}c@{\hskip 0.5in}c@{\hskip 0.5in}c}
\hline
\hline
Formula for $\alpha$-decay half-life                       & RMSE  &$\overline{\delta}$&F & No. of Coefficients \\
\hline
M Horoi (2020) Set-1         &    0.8355& 	0.5752  &3.7597   &  24  \\
M Horoi (2020) Set-3         &    0.9383&	0.6345	&4.3098    &  9 \\
M Sobic. (2020)              &    0.9833&	0.6416	&4.3813    &  24\\
Sobiczewski (2005)           &    1.1788&	0.5821	&3.8203    &  7 \\
Akre (2017)                  &    1.1813&	0.6038	&4.0162    &  20\\
SPm1 (2009)                  &    1.2242&	0.6964	&4.9699    &  7 \\
VSm1 (2009)                  &    1.2835&	0.7403	&5.4988    &  8 \\
NUDL (2018)                  &    1.2860&	0.6721	&4.7004    &  4 \\
MSLB (2019)                  &    1.2978&	0.7111	&5.1411    &  16\\
Rm (2009)                    &    1.3119&	0.7220	&5.2722    &  12\\
NSobicP (2018)               &    1.3416&	0.7779	&5.9967    &  4 \\
New RenA (2019)              &    1.3462&	0.6912	&4.9116    &  20\\
NRDX (2008)                  &    1.3469&	0.8347	&6.8345    &  3 \\
MVS (2019)                   &    1.3646&	0.6931	&4.9331    &  24\\
Universal curve (2011)       &    1.4081&	0.7932	&6.2112    &  5 \\
MUDL (2019)                  &    1.4357&	0.8168	&6.5581    &  20\\
N Ro (2018)                  &    1.4426&	0.9017	&7.9752    &  4 \\
mB1 (2016)                   &    1.5392&	0.9973	&9.9387    &  7 \\
SemFIS2 (2010)               &    1.6678&	1.0076	&10.1755   &  11\\
mB2 (2016)                   &    1.8361&	1.0918	&12.3547   &  12\\
VSm2 (2009)                  &    1.9839&	1.0858	&12.1847   &  16\\
Horoi (2004)                 &    2.5237&	2.3267	&212.1864  &  5 \\
Manjunatha (2019)            &    2.6752&	1.5039	&31.9113   &  4 \\
SPm2 (2009)                  &    2.8200&	1.4614	&28.9302   &  16\\
\hline
\hline
\end{tabular}}
\label{rmse-alpha}
\end{table}

\begin{figure}[!htbp]
\centering
\includegraphics[width=0.9\textwidth]{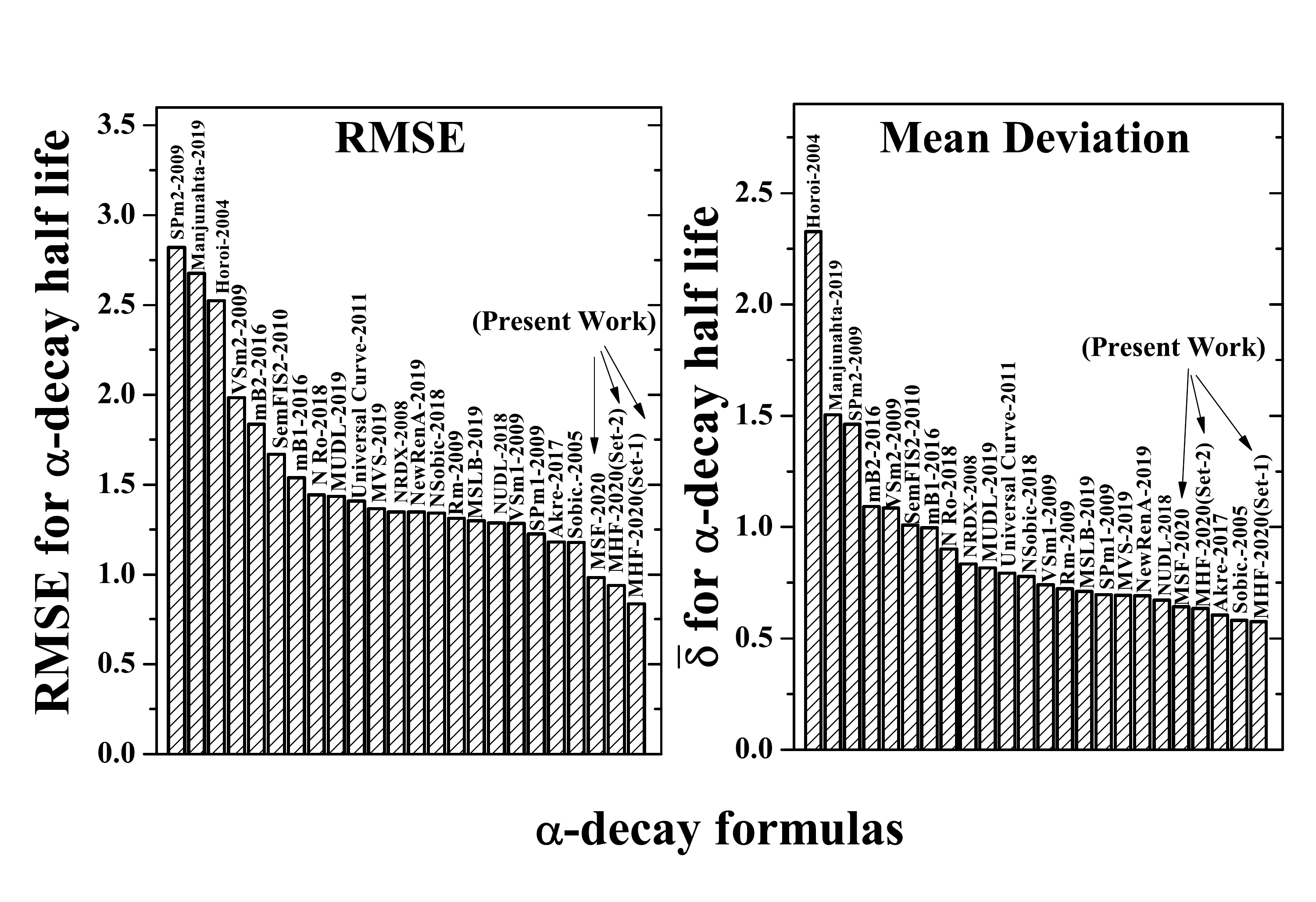}
\caption{Root mean square error (RMSE) and mean deviation ($\overline{\delta}$) of various formulas for $\alpha$-decay half-lives for the nuclei in the range of 82$\leq$Z$\leq$118.}\label{fig3}
\end{figure}

The root mean square error (RMSE), mean deviation ($\overline{\delta}$) and the index F$=$10$^{\overline{\delta}}$ expressed by the following equations are deduced for all the above mentioned formulas and mentioned in Table \ref{rmse-alpha}. The total number of fitted coefficients are also listed in front of each formulas. A justified comparison among formulas with coefficients $>$10 leads the modified formulas with a good merit. Also, the RMSE and $\overline{\delta}$ are plotted in Fig. \ref{fig3} to depict a comparison of the above mentioned formulas. It is worth to mention here that few formulas with comparatively lesser number of coefficients (Sobiczewski, NUDL etc.) still possess a good accuracy for the calculation of half lives.

\begin{eqnarray}
RMSE &=& \sqrt{\frac{1}{N_{nucl}}\sum^{N_{nucl}}_{i=1}\left(log\frac{T^i_{th}}{T^i_{exp}}\right)^2}\\
\overline{\delta} &=& \frac{1}{N_{nucl}}\sum^{N_{nucl}}_{i=1}\left|log\frac{T^i_{th}}{T^i_{exp}}\right|\\
F &=& 10^{\overline{\delta}}
\end{eqnarray}

It is gratifying to note from Table \ref{rmse-alpha} and Fig. \ref{fig3} that the modified formulas introduced in present work, result with minimum error and deviation comparative to other 21 considered formulas. In fact, comparison shown in the Table \ref{rmse-alpha} for all the statistical parameters certifies the applicability of the MHF-2020 (set-1 and set-2) and MSF-2020, and therefore will be employed for prediction of decay modes, to calculate $\alpha$-decay half-lives and to construct decay chains in the unknown territory of superheavy land.\par

\begin{table}[!htbp]
\caption{Root mean square error (RMSE), mean deviation ($\overline{\delta}$), index F$=$10$^{\overline{\delta}}$ (in the unit of hour) and No. of coefficients of various formulas for SF half-lives calculated for nuclei in the range with 82$\leq$Z$\leq$118.}
\centering
\resizebox{1.0\textwidth}{!}{%
\begin{tabular}{l@{\hskip 0.5in}c@{\hskip 0.5in}c@{\hskip 0.5in}c@{\hskip 0.5in}c}
\hline
\hline
Formula for SF half-life                                              & RMSE  &$\overline{\delta}$&F& No. of Coefficients \\
\hline
Modified Bao (MBF-2020)       & 2.1103    & 	1.7389&	0.0152  &  7  \\
Bao (2015)                    & 2.8379    & 	2.3107&	0.0568  &  8  \\
Xu (2008)                     & 3.1387    & 	2.5305&	0.0942  &  7  \\
Santosh (2010)                & 3.3064    & 	2.7297&	0.1491  &  5  \\
Karpov (2012)                 & 5.6784    & 	5.1697&	41.0536 &  9  \\
Ren (2005)                    & 5.8511    & 	4.0182&	2.8967  &  7  \\
Silisteanu (2015)             & 6.0458    & 	5.6374&	120.5425&  10 \\
\hline
\hline
\end{tabular}}
\label{SF-rmse}
\end{table}

\begin{figure}[!htbp]
\centering
\includegraphics[width=0.7\textwidth]{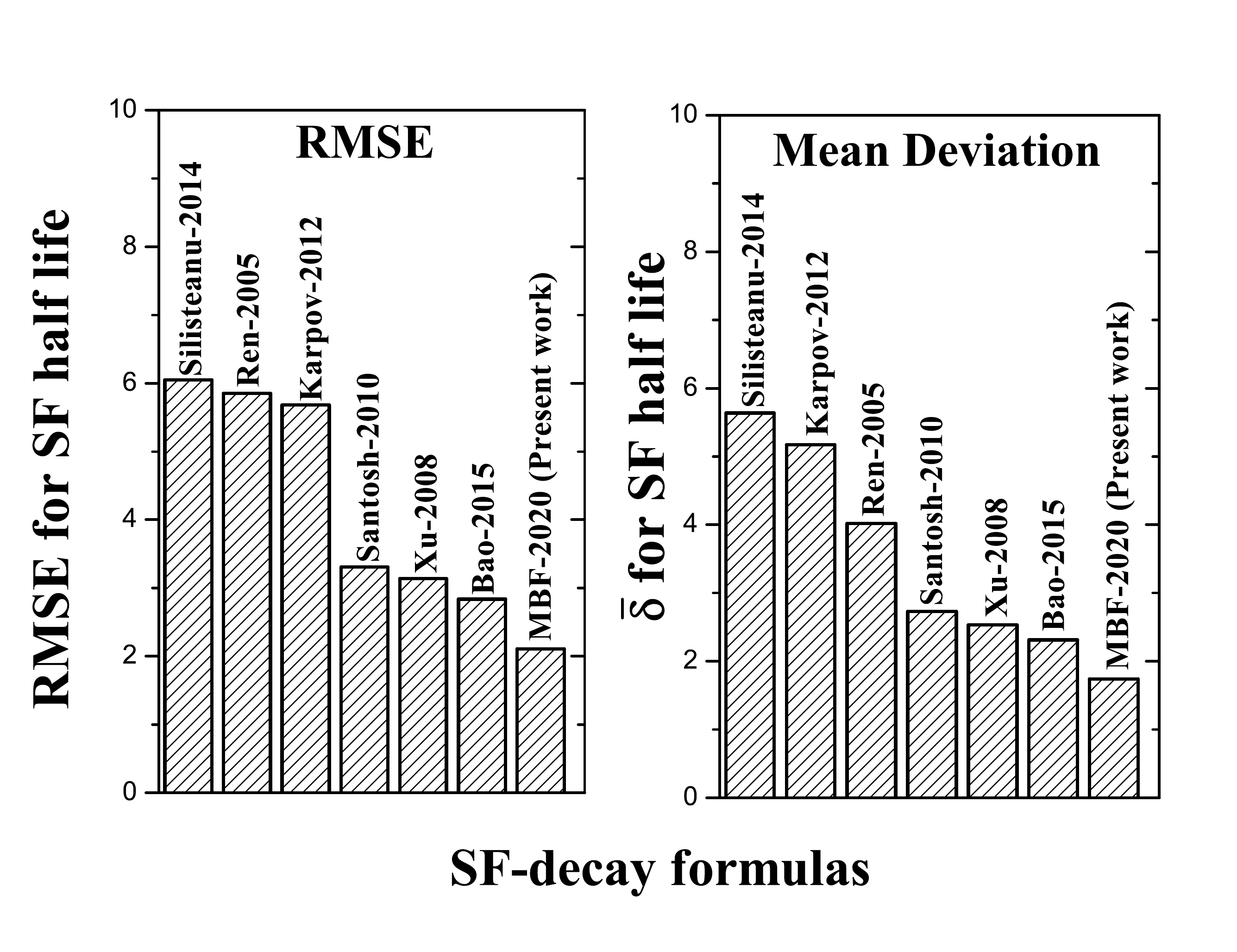}
\caption{Root mean square error (RMSE) and mean deviation ($\overline{\delta}$) of various formulas for SF half-lives for the nuclei in the range of 82$\leq$Z$\leq$118.}\label{fig4}
\end{figure}

In a similar manner, to find more precise formula for spontaneous fission: the mode which compete with $\alpha$-decay, we apply the formula given by Ren \textit{et al.} \cite{ren2005}, Xu \textit{et al.} \cite{xu2008}, Santosh \textit{et al.} \cite{santosh2010}, Karpov \textit{et al.} \cite{karpov2012}, Silisteanu \textit{et al.} \cite{silisteanu2015}, Bao \textit{et al.} \cite{bao2015} along with the modified Bao formula of present work. Spontaneous fission half-lives are calculated by using above all formulas for 49 nuclei for which SF-half-lives are known and consequently a few statistical parameters are obtained which are listed in Table \ref{SF-rmse} and also shown in Fig. \ref{fig4}.  The comparison displays improvement in the prediction of SF half-lives after the modification in the formula of Bao \textit{et al.} \cite{bao2015} which earmark the use of modified Bao formula (MBF-2020) to analyze the competition between $\alpha$-decay and SF.\par

              As mentioned above, four methods of machine learning i.e. Decision Trees (DTs), Random Forest (RF), Multilayer Perceptron (MLP) neural network, and XGBoost are tested for $\alpha$-decay and SF half-lives considering 366 and 49 nuclei, respectively. We apply 5-fold cross-validation, and accordingly, the dataset is split into 5 groups. Using this 5-fold cross-validation, root mean square error (RMSE) and mean deviation ($\overline{\delta}$) of considered methods of machine learning are mentioned in Table \ref{ML-RMSE}.

\begin{table}[!htbp]
\caption{Root mean square error (RMSE) and mean deviation ($\overline{\delta}$) of various methods of machine learning for the $\alpha$-decay half-lives and SF half-lives data in the range with 82$\leq$Z$\leq$118.}
\centering
\resizebox{0.7\textwidth}{!}{%
\begin{tabular}{l@{\hskip 0.5in}c@{\hskip 0.5in}cc@{\hskip 0.5in}c@{\hskip 0.5in}c}
\hline
\hline
\multicolumn{1}{l}{Machine learning}&\multicolumn{2}{c}{RMSE}&\multicolumn{1}{c}{}&\multicolumn{2}{c}{$\overline{\delta}$}\\
\cline{2-3} \cline{5-6}
method&   $\alpha$-decay & SF  &&   $\alpha$-decay & SF \\
\hline
MLP neural network          & 0.9774  & 2.5753 && 0.6459  & 2.5177  \\
XGBoost                     & 1.0245  & 3.1719 && 0.6468  & 2.6695   \\
Random Forest               & 1.1134  & 3.2748 && 0.8035  & 2.9100   \\
Decision Trees (DTs)        & 1.4449  & 4.0971 && 1.0780  & 3.6270   \\
\hline
\hline
\end{tabular}}
\label{ML-RMSE}
\end{table}

The Table \ref{ML-RMSE} establishes MLP neural network and XGBoost methods as the more appropriate method of machine learning among the considered ones because for these two methods the RMSE and $\overline{\delta}$ for $\alpha$-decay and SF half-lives are found very similar to what are found for modified empirical formulas (please see Tables \ref{rmse-alpha} and \ref{SF-rmse}). Therefore, in the following, we will employ MLP and XGBoost for the prediction of $\alpha$-decay half-lives. For SF half-lives, since the training data is very limited therefore predictions of MLP neural network for such a small dataset are not found reasonable. As a result, we opt the Random Forest (RF) method to predict SF half-lives along with XGBoost algorithm.\par

With the above-mentioned selection of empirical formulas of $\alpha$-decay and SF half-lives along with machine learning methods, in Table \ref{prediction}, we predict the decay modes of nuclei in the range 112$\leq$Z$\leq$118 to demonstrate the predictive power of modified formulas and machine learning methods. It is important to mention here that in all the formulas of $\alpha$-decay considered here (3 Proposed$+$21 others), the contribution of centrifugal potential is not included. The calculations show that favored $\alpha$-decay in which the orbital angular momentum taken by the $\alpha$-particle is zero ($l$$=$0), can be efficaciously described by the considered formulas. However, for unfavored $\alpha$-decay ($l$$\neq$0) one needs to take into account the $l$ dependence which is left for our upcoming work.

In Table \ref{prediction}, a few nuclei from 112$\leq$Z$\leq$118 are taken for which change in angular momentum $l$$=$0 (favoured) and the values of Q$_{\alpha}$ are evaluated from latest database \cite{nndc}. Few available $log_{10}$T$_{1/2}$ and decay modes are also tabulated for comparison. To predict $\alpha$-decay half-lives we use modified Horoi formula (MHF-2020) (set-1 and set-2) and modified Sobiczewski formula (MSF-2020) which are constructed in this present work. Also, to predict SF half-lives we employ modified Bao formula (MBF-2020) reported in this work. These formulas are utilized to analyze the competition of $\alpha$-decay with SF and consequently to predict probable decay mode. The same kind of predictions are also described in Table \ref{prediction} by using machine learning methods as an alternate approach.\par

\begin{table}[htbp]
\caption{Prediction of decay modes and half-lives from modified formulas and machine learning for the range 112$\leq$Z$\leq$118.}
\centering
\resizebox{1.00\textwidth}{!}{%
\begin{tabular}{cc|cccc|ccccc|cccccc}
\hline
\hline
\multicolumn{1}{c}{Z}&
\multicolumn{1}{c|}{A}&
\multicolumn{4}{c|}{Experimental Data \cite{nndc}}&
\multicolumn{5}{c|}{Modified formulas of present work}&
\multicolumn{6}{c}{Machine learning methods}\\
\hline
\multicolumn{1}{c}{}&
\multicolumn{1}{c|}{}&
\multicolumn{1}{c}{Q$_{\alpha}$}& \multicolumn{1} {c} {l} & \multicolumn{1}{c}{log$_{10}$T$_{1/2}$(s)}& \multicolumn{1}{c|}{Decay}&
\multicolumn{3}{c}{log$_{10}$T$_{1/2}^{\alpha}$(s)}& \multicolumn{1}{c}{log$_{10}$T$_{1/2}^{SF}$(s)}& \multicolumn{1}{c|}{Predicted}&
\multicolumn{2}{c}{log$_{10}$T$_{1/2}^{\alpha}$(s)}& \multicolumn{1}{c}{} &\multicolumn{2}{c}{log$_{10}$T$_{1/2}^{SF}$(s)}& \multicolumn{1}{c}{Predicted}\\
\cline{7-9}  \cline{12-13}   \cline{15-16}
\multicolumn{1}{c}{}&
\multicolumn{1}{c|}{}&
\multicolumn{1}{c}{(MeV)}&\multicolumn{1}{c}{}&\multicolumn{1}{c}{}&\multicolumn{1}{c|}{modes}&
\multicolumn{1}{c}{MHF (Set-1)}& \multicolumn{1}{c}{MHF (Set-2)}& \multicolumn{1}{c}{MSF}& \multicolumn{1}{c}{MBF}& \multicolumn{1}{c|}{Decay-modes}&
\multicolumn{1}{c}{MLP}& \multicolumn{1}{c}{Xgboost}&\multicolumn{1}{c}{} & \multicolumn{1}{c}{RF}& \multicolumn{1}{c}{Xgboost}& \multicolumn{1}{c}{Decay-modes}\\
\hline

112 & 276 & 11.9 & 0 &        &              & -6.27  & -5.32  &   -3.84  &  3.14   &  $\alpha$  &  -3.54  &   -4.87 && 0.29   & -0.61     & $\alpha$    \\

112 & 278 & 11.3 & 0 &        &              & -4.78  & -3.92  &   -2.64  &  -1.79  &  $\alpha$  &  -2.52  &   -3.19 && 0.21   & -0.65     & $\alpha$    \\

112 & 280 & 10.7 & 0 &        &              & -3.18  & -2.40  &   -1.36  &  -5.10  &  SF        &  -1.19  &   -1.75 && 0.00   & -0.75     & $\alpha$    \\

112 & 282 & 10.2 & 0& -3.30  &   SF         & -1.48  & -0.78  &   -0.01  &  -4.26  &  SF        &  -0.51  &   -0.31 && -0.02  & -0.68      & SF    \\

112 & 284 & 9.6  & 0& -1.00  &   SF         & 0.44   & 1.03  &  1.48     &  -2.49  &  SF        &  0.73   &   1.15  && 0.05   & -0.49     & SF    \\


114 & 284 & 10.8 & 0&-2.60   &   SF         & -2.67  & -1.88  &   -1.10  &  0.13   &  $\alpha$  &  -1.32  &   -1.28 && -0.33  & -1.24      &  $\alpha$    \\
114 & 285 & 10.6 & 0&-0.82   &   $\alpha$   & -1.07  & -0.78  &   -1.03  &  3.48   &  $\alpha$  &  -0.95  &   -0.66 && -0.22  & -0.92      &  $\alpha$/SF   \\
114 & 286 & 10.4 & 0 &-0.80   &   SF/$\alpha$& -1.36  & -0.64  &   -0.03  &  1.16   &  $\alpha$  &  -0.96  &   -0.18 && -0.16  & -0.99      &  SF/$\alpha$    \\
114 & 287 & 10.2 & 0&-0.29   &   $\alpha$   & 0.12   & 0.41  &  0.07     &  4.64   &  $\alpha$  &  -0.04  &   0.36  && -0.16  & -0.99      &  SF    \\
114 & 288 & 10.1 & 0 &-0.28   &   $\alpha$   & -0.43  & 0.25  &   0.80    &  2.42   &  $\alpha$  &  0.02   &   0.56  && -0.16  & -0.99      &  SF    \\
114 & 289 & 10.0 & 0&-0.01   &   $\alpha$   & 0.70   & 0.98  &  0.63     &  6.00   &  $\alpha$  &  0.05   &   0.80  && -0.16  & -0.99      &  SF    \\

116 & 290 & 11.0 & 0 &-1.82   &   $\alpha$   & -2.67  & -1.81  &   -0.94  &  4.41   &  $\alpha$  &  -1.44  &   -1.20 && -0.16  & -0.99      &  $\alpha$    \\

116 & 292 &10.8 &  0 &-1.74   &   $\alpha$   & -2.04  & -1.20  &   -0.31  &  4.81   &  $\alpha$  &  -1.46  &   -0.65 && -0.16  & -0.99      &  $\alpha$    \\

118 & 294 &11.6 & 0& -3.05   &   $\alpha$   & -3.80  &-2.82	&   -1.72  &  6.15   &  $\alpha$  &  -1.59	&   -0.40 && -0.16  & -0.99      &  $\alpha$    \\
\hline
\hline
\end{tabular}}
\label{prediction}
\end{table}

\begin{figure}[htbp]
\centering
\includegraphics[width=0.7\textwidth]{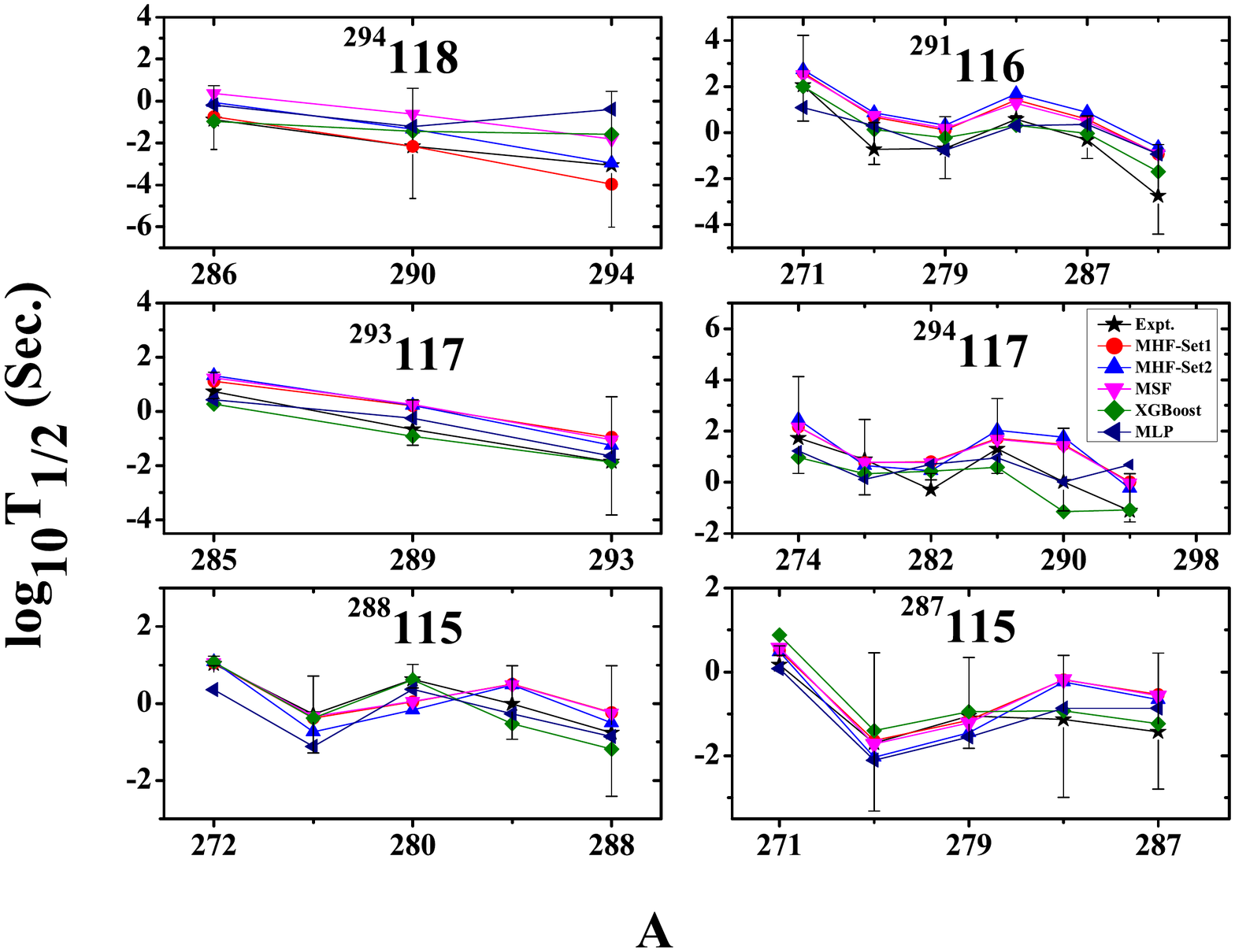}
\caption{(Colour online) $\alpha$ half-lives for the experimentally known decay chains \cite{Oganessian2010,ogan2006,ogan2004,ogan2012}: $^{294}$Og, $^{294}$Ts, $^{293}$Ts, $^{291}$Lv, $^{288}$Mc and $^{287}$Mc.}\label{fig5}
\end{figure}

Table \ref{prediction} itself expresses the predictive power of modified formulas and also the use of machine learning methods as the decay modes are anticipated effectively. In fact, not only the decay modes are found in an excellent match with available experimental/estimated decay modes but also the half-lives are found in agreement with experimental half-lives. In addition, we have compared our estimation of half-lives for the experimentally known decay chains \cite{Oganessian2010,ogan2006,ogan2004,ogan2012}: $^{294}$Og, $^{294}$Ts, $^{293}$Ts, $^{291}$Lv, $^{288}$Mc and $^{287}$Mc in Fig. \ref{fig5}. The $\alpha$-decay half-lives are calculated by employing MHF-2020 (set-1 and set-2) and MSF-2020 which are constructed in this present work along with machine learning methods: XGBoost and MLP. Satisfyingly, these half-lives are found in a reasonable match with the experimental decay half-lives mentioned in Fig. \ref{fig5} with error bars.

This agreement countenances the use of formulas and methods reported in the present work to estimate decay modes, decay chains, and half-lives for the unknown territory of superheavy nuclei to accompaniment the crucial experiments eyeing on heavier elements with Z$\geq$118 \cite{hofmann2016,Oganessian2009,voinov2020,brewer2018}. With this in view, we employ the above-mentioned formulas and machine learning methods to construct the possible decay chains for isotopes of Og and Z$=$119 which are of current interest. The decay chains are constructed for neutron number lying between 168$\leq$N$\leq$184 as observed from experimental decay chains \cite{Oganessian2010,ogan2006,ogan2004,ogan2012} and speculated for island of stability at N$\sim$184 \cite{hofmann2016,Oganessian2009,voinov2020,brewer2018}. To apply empirical formulas for calculations of $\alpha$-decay half-lives the experimental and evaulated Q$_{\alpha}$ \cite{nndc} are used. For the rest of the nuclei, the values of Q$_{\alpha}$ can be taken from theoretical approaches. In this regard, we have performed calculations using relativistic mean-field theory \cite{saxena1,saxena2,saxena3,saxena4} for the nuclei in the range of 111$\leq$Z$\leq$118. The values of Q$_{\alpha}$ are also taken directly from FRDM \cite{moller2019}, RCHB \cite{rchb2018} and KTUY \cite{ktuy2005} mass tables for comparison. For determination of Q$_{\alpha}$, one can also use the very efficient five-parameter formula derived from LDM considerations by Dong \textit{et al.} \cite{dongprc2010,dong2011}, which is given by:
\begin{eqnarray}\label{dong-q-alpha}
Q_{\alpha} (MeV) = aZA^{-4/3}(3A-Z)+b \left(\frac{N-Z}{A}\right)^2 \nonumber\\
+ c \left[\frac{|N-152|}{N}-\frac{|N-154|}{N-2}\right] \nonumber\\
+ d \left[\frac{|Z-110|}{N}-\frac{|Z-112|}{Z-2}\right] + e
\end{eqnarray}

\begin{figure}[htbp]
\centering
\includegraphics[width=0.54\textwidth]{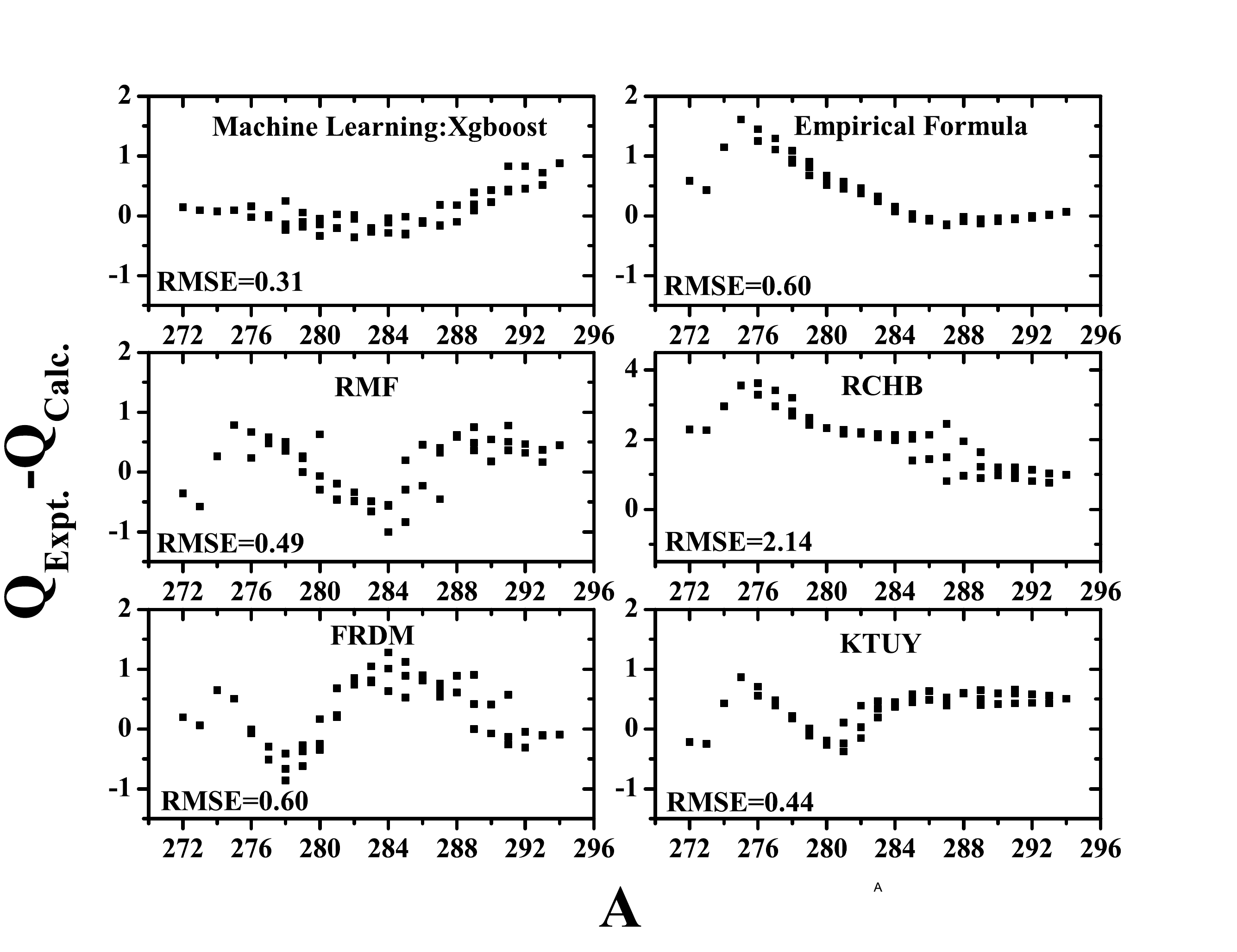}
\caption{Errors in the values of Q$_{\alpha}$ for the nuclei in range 111$\leq$Z$\leq$118.}\label{fig6}
\end{figure}

In addition to the above treatments for Q$_{\alpha}$, we have applied machine learning methods as an alternate path. To train a machine, we have used all experimental/evaulated Q$_{\alpha}$ values ($\approx$709) between 82$\leq$Z$\leq$110 with the 5-fold cross-validation as described above. Once again XGBoost is found with minimum RMSE among the considered machine learning methods and, therefore, we have predicted Q$_{\alpha}$ values for the nuclei in the range 111$\leq$Z$\leq$118 using XGBoost algorithm. The errors in the values of Q$_{\alpha}$ for the nuclei in the range 111$\leq$Z$\leq$118 calculated by RMF \cite{saxena1,saxena2,saxena3,saxena4}, FRDM \cite{moller2019}, RCHB \cite{rchb2018}, KTUY \cite{ktuy2005}, Dong \textit{et al.} formula \cite{dongprc2010,dong2011} and XGBoost are plotted in Fig. \ref{fig6}. RMSE for all these approaches are also mentioned in the corresponding panels.

It is clear from Fig. \ref{fig6} and mentioned RMSE values that XGBoost is found closer with experimental/evaulated Q$_{\alpha}$ values comparative to the theoretical approaches and the formula. Hence, for the prediction of half-lives for Og isotopes and Z$=$119 isotopes, we will opt Q$_{\alpha}$ values from XGBoost algorithm where experimental Q$_{\alpha}$ are not available. In this way, with the more accurate Q$_{\alpha}$, we expect a more reliable prediction of $\alpha$-decay half-lives using more precise formulas (MHF-2020 \& MSF-2020, please see Table \ref{rmse-alpha}) reported in the present work. In Tables \ref{prediction1} \& \ref{prediction2}, we show decay chains of Og isotopes in the range 286$\leq$A$\leq$302. As mentioned above, to speculate chances of favoured and unfavoured $\alpha$-decay, we have also mentioned change in the angular momentum ($l$) calculated from selection rule \cite{denisov2009,royer2020} based on parity and spin of parent and daughter nuclei taken from latest evaluated nuclear properties table NUBASE2016 \cite{audi2017} or P. M\"{o}ller \cite{mollerparity}. More detailed analysis on favoured and unfavoured $\alpha$-decay will be covered in our upcoming work. \par

Asterisk (*) values of Q$_{\alpha}$ are taken from the XGBoost algorithm whereas all other values are experimental/evaluated \cite{nndc}. Using these Q$_{\alpha}$ values, $\alpha$-decay half-lives ($log_{10}$T$_{1/2}^{\alpha}$(s)) are deduced by modified Horoi formula (MHF-2020) (set-1 and set-2) and modified Sobiczewski formula (MSF-2020), and also with MLP and XGBoost algorithm of machine learning. Spontaneous fission half-lives ($log_{10}$T$_{1/2}^{SF}$(s)) are also mentioned which are calculated by using modified Bao formula (MBF-2020). Consequently, decay modes are predicted for all the chains which are compared with available experimental/estimated decay modes \cite{nndc}. Similar calculations and prediction are tabulated for Z$=$119 isotopes in the range 287$\leq$A$\leq$303 in Tables \ref{prediction3} \& \ref{prediction4}. \par

From Tables \ref{prediction1} to \ref{prediction4}, it is indulging to note that our prediction of decay modes and half-lives are in an excellent match with available experimental data. $^{286-291}$Og and $^{287-296,302,303}$119 are found with very long $\alpha$-decay chains (6$\alpha$/7$\alpha$) and are found as potential candidates for future experiments. As far as the current experimental facilities are concerned, $^{50}$Ti+$^{249}$Bk and $^{50}$Ti+$^{249-251}$Cf reactions could lead to the discovery of $^{295,296}$119 and $^{291}$Og \cite{voinov2020} which are found with 7$\alpha$ decay chains as per our predictions. For $^{48}$Ca beam, the available heaviest target material is $^{251}$Cf (N = 153) which could lead to the production of $^{295}$Og, $^{296}$Og, and $^{297}$Og \cite{Oganessian2015} and from our prediction, these isotopes of Og are indeed found with 3$\alpha$/4$\alpha$ chain and their half-lives are found within the range of current experimental facilities. Therefore, our predictions from the modified formulas are expected to provide more extensive and precise information for the upcoming experiments to reach the boundaries of the island of stability.
\begin{table}[htbp]
\caption{Prediction of decay chains from modified formulas and machine learning for Og isotopes. Q$_{\alpha}$  and log$_{10}$T$_{1/2}$ are taken from Ref. \cite{nndc}, whereas (*) values of Q$_{\alpha}$ are taken from XGBoost algorithm as described in text. The angular momentum ($l$) is calculated from selection rule \cite{denisov2009,royer2020} based on parity and spin of parent and daughter nuclei, which are taken from latest evaluated nuclear properties table NUBASE2016 \cite{audi2017} or P. M\"{o}ller \cite{mollerparity}.}
\centering
\resizebox{0.85\textwidth}{!}{%
\begin{tabular}{c|ccc|ccccc|c|cc}
\hline
\hline
\multicolumn{1}{c|}{Nucleus}&
\multicolumn{3}{c|}{Expt.}&
\multicolumn{5}{c|}{log$_{10}$T$_{1/2}^{\alpha}$(s)}&
\multicolumn{1}{c|}{log$_{10}$T$_{1/2}^{SF}$(s)}&
\multicolumn{2}{c}{Decay Modes}\\
\cline{2-11}
\multicolumn{1}{c|}{}&
\multicolumn{1}{c}{Q$_{\alpha}$}&
\multicolumn{1}{c}{l}&
\multicolumn{1}{c|}{log$_{10}$T$_{1/2}$(s)}&
\multicolumn{1}{c}{MHF (Set-1)}&
\multicolumn{1}{c}{MHF (Set-2)}&
\multicolumn{1}{c}{MSF}&
\multicolumn{1}{c}{XGBoost}&
\multicolumn{1}{c|}{MLP}&
\multicolumn{1}{c|}{MBF}&
\multicolumn{1}{c}{Predicted}&
\multicolumn{1}{c}{Expt.}\\
\hline
$^{286}$Og & 11.17*  & 0 &          &  -2.19    &   -1.39   &  -1.37  &  -1.07    &  -2.83   &  1.59    &  $\alpha$  &             \\
$^{282}$Lv & 11.18*  & 0 &         &  -2.92    &   -2.12   &  -1.83  &  -1.71    &  -2.97   &  -0.45   &  $\alpha$  &             \\
$^{278}$Fl & 11.35*  & 0 &         &  -4.06    &   -3.23   &  -2.57  &  -2.74    &  -3.21   &  -2.71   &  $\alpha$  &             \\
$^{274}$Cn & 10.84*  & 0 &         &  -3.21    &   -2.50   &  -1.99  &  -1.97    &  -2.72   &  3.01    &  $\alpha$  &             \\
$^{270}$Ds & 11.12   & 0  & -4.00   &  -4.66    &   -3.92   &  -2.97  &  -3.34    &  -3.33   &  0.21    &  $\alpha$  &  $\alpha$   \\
$^{266}$Hs & 10.35   & 0  & -2.64   &  -3.09    &   -2.50   &  -1.87  &  -1.91    &  -2.29   &  -1.39   &  $\alpha$  &  $\alpha$   \\
$^{262}$Sg & 9.60    & 0  & -2.16   &  -1.48    &   -1.04   &  -0.71  &  -0.55    &  -0.29   &  -2.09   &  SF        &  SF         \\
\hline
$^{287}$Og & 11.15*  & 0 &         &  -1.31    &   -0.94   &  -1.45  &  -0.96    &  -2.77   &  5.05    &  $\alpha$  &             \\
$^{283}$Lv & 11.09*  & 2 &         &  -1.80    &   -1.45   &  -1.84  &  -1.39    &  -2.84   &  3.33    &  $\alpha$  &             \\
$^{279}$Fl & 11.29*  & 5 &         &  -2.99    &   -2.67   &  -2.87  &  -2.53    &  -2.98   &  0.31    &  $\alpha$  &             \\
$^{275}$Cn & 11.12*  & 5 &         &  -3.18    &   -2.88   &  -3.01  &  -2.69    &  -2.78   &  6.14    &  $\alpha$  &             \\
$^{271}$Ds & 10.87   & 2 & -2.79   &  -3.17    &   -2.88   &  -2.98  &  -2.66    &  -2.92   &  5.42    &  $\alpha$  &  $\alpha$   \\
$^{267}$Hs & 10.04   & 4 & -1.28   &  -1.50    &   -1.23   &  -1.44  &  -1.09    &  -1.11   &  3.11    &  $\alpha$  &  $\alpha$   \\
$^{263}$Sg & 9.40    & 2 & 0.00    &  -0.25    &   -0.03   &  -0.27  &  -0.02    &  0.29    &  0.27    &  $\alpha$  &  $\alpha$   \\
\hline
$^{288}$Og & 11.12*  & 0 &         &  -2.06    &   -1.28   &  -1.16  &  -0.81    &  -2.62   &  3.17    &  $\alpha$  &             \\
$^{284}$Lv & 10.99*  & 0 &         &  -2.38    &   -1.61   &  -1.34  &  -1.09    &  -2.51   &  1.44    &  $\alpha$  &             \\
$^{280}$Fl & 11.14*  & 0 &         &  -3.53    &   -2.73   &  -2.06  &  -2.12    &  -2.81   &  -1.40   &  $\alpha$  &             \\
$^{276}$Cn & 11.90   & 0  &         &  -6.27    &   -5.32   &  -3.84  &  -4.87    &  -3.54   &  3.14    &  $\alpha$  &             \\
$^{272}$Ds & 10.80   & 0  &         &  -3.86    &   -3.14   &  -2.22  &  -2.49    &  -2.40   &  3.93    &  $\alpha$  &            \\
$^{268}$Hs & 9.62    & 0  & -0.40   &  -0.89    &   -0.42   &  -0.16  &  0.01     &  -0.38   &  1.57    &  $\alpha$  &  $\alpha$   \\
$^{264}$Sg & 9.21    & 0  & -1.43   &  -0.26    &   0.13    &  0.35   &  0.44     &  0.93    &  -1.10   &  SF        &  SF         \\
\hline
$^{289}$Og & 11.09*  & 0 &         &  -1.17    &   -0.81   &  -1.30  &  -0.71    &  -2.59   &  6.65    &  $\alpha$  &             \\
$^{285}$Lv & 10.99*  & 2 &         &  -1.56    &   -1.23   &  -1.60  &  -1.07    &  -2.40   &  4.69    &  $\alpha$  &             \\
$^{281}$Fl & 10.96*  & 2 &         &  -2.14    &   -1.83   &  -2.07  &  -1.63    &  -2.41   &  1.41    &  $\alpha$  &             \\
$^{277}$Cn & 11.62   & 5 &         &  -4.53    &   -4.26   &  -4.16  &  -4.03    &  -2.86   &  3.36    &  $\alpha$  &             \\
$^{273}$Ds & 11.37   & 5 & -3.77   &  -4.53    &   -4.26   &  -4.13  &  -4.00    &  -3.11   &  6.83    &  $\alpha$  &  $\alpha$   \\
$^{269}$Hs & 9.34    & 2 & 0.99    &  0.66     &   0.89    &  0.55   &  0.72     &  0.66    &  5.11    &  $\alpha$  &  $\alpha$   \\
$^{265}$Sg & 9.05    & 4 & 1.16    &  0.87     &   1.05    &  0.80   &  0.86     &  1.08    &  2.47    &  $\alpha$  &  $\alpha$   \\
\hline
$^{290}$Og & 11.05*  & 0 &         &  -1.92    &   -1.13   &  -0.92  &  -0.59    &  -2.32   &  4.50    &  $\alpha$  &             \\
$^{286}$Lv & 10.92*  & 0 &         &  -2.23    &   -1.46   &  -1.07  &  -0.89    &  -2.21   &  2.97    &  $\alpha$  &             \\
$^{282}$Fl & 10.69*  & 0 &         &  -2.23    &   -1.51   &  -1.03  &  -0.94    &  -1.69   &  -0.95   &  $\alpha$  &             \\
$^{278}$Cn & 11.31   & 0  &         &  -4.78    &   -3.92   &  -2.64  &  -3.19    &  -2.52   &  -1.79   &  $\alpha$  &             \\
$^{274}$Ds & 11.70   & 0  &         &  -6.47    &   -5.54   &  -3.73  &  -5.05    &  -3.01   &  4.96    &  $\alpha$  &             \\
$^{270}$Hs & 9.07    & 0  & 1.34    &  0.98     &   1.36    &  1.33   &  1.40     &  0.84    &  4.52    &  $\alpha$  &  $\alpha$   \\
$^{266}$Sg & 8.80    & 0  & 1.32    &  1.13     &   1.47    &  1.55   &  1.48     &  1.31    &  0.62    &  SF        &  SF         \\
\hline
$^{291}$Og & 11.07*  & 2 &         &  -1.15    &   -0.80   &  -1.24  &  -0.66    &  -2.24   &  7.93    &  $\alpha$  &             \\
$^{287}$Lv & 10.82*  & 0 &         &  -1.08    &   -0.76   &  -1.14  &  -0.63    &  -2.07   &  5.86    &  $\alpha$  &             \\
$^{283}$Fl & 10.59*  & 4 &         &  -1.11    &   -0.81   &  -1.11  &  -0.68    &  -1.62   &  2.21    &  $\alpha$  &             \\
$^{279}$Cn & 11.04   & 4 &         &  -3.07    &   -2.81   &  -2.81  &  -2.52    &  -1.69   &  -0.76   &  $\alpha$  &             \\
$^{275}$Ds & 11.40   & 0 &         &  -4.67    &   -4.43   &  -4.18  &  -4.18    &  -2.45   &  6.15    &  $\alpha$  &             \\
$^{271}$Hs & 9.51    & 0 &         &  0.06     &   0.28    &  0.07   &  0.22     &  0.41    &  7.21    &  $\alpha$  &             \\
$^{267}$Sg & 8.63    & 0 &         &  2.32     &   2.48    &  2.16   &  1.93     &  1.53    &  4.90    &  $\alpha$  &             \\
\hline
$^{292}$Og & 11.02*  & 0 &         &  -1.90    &   -1.10   &  -0.74  &  -0.55    &  -1.99   &  6.14    &  $\alpha$  &             \\
$^{288}$Lv & 10.77*  & 0 &         &  -1.82    &   -1.06   &  -0.62  &  -0.53    &  -1.79   &  3.40    &  $\alpha$  &             \\
$^{284}$Fl & 10.80   & 0 & -2.60   &  -2.67    &   -1.88   &  -1.10  &  -1.28    &  -1.32   &  0.13    &  $\alpha$  &  SF         \\
$^{280}$Cn & 10.73   & 0  &         &  -3.18    &   -2.40   &  -1.36  &  -1.75    &  -1.19   &  -5.10   &  SF        &             \\
$^{276}$Ds & 11.11   & 0  &         &  -5.01    &   -4.14   &  -2.49  &  -3.39    &  -2.17   &  2.08    &  $\alpha$  &             \\
$^{272}$Hs & 9.78    & 0  &         &  -1.63    &   -1.03   &  -0.17  &  -0.53    &  0.08    &  3.58    &  $\alpha$  &             \\
$^{268}$Sg & 8.30    & 0  &         &  3.01     &   3.28    &  3.08   &  2.97     &  2.14    &  2.98    &  $\alpha$  &             \\
\hline
$^{293}$Og & 11.01*  & 2 &          &  -0.99    &   -0.66   &  -1.06  &  -0.54    &  -2.08   &  9.89    &  $\alpha$  &             \\
$^{289}$Lv & 11.10   & 2 &          &  -1.94    &   -1.65   &  -1.84  &  -1.44    &  -1.58   &  6.64    &  $\alpha$  &             \\
$^{285}$Fl & 10.56   & 0 &  -0.82   &  -1.07    &   -0.78   &  -1.03  &  -0.66    &  -0.95   &  3.48    &  $\alpha$  &  $\alpha$   \\
$^{281}$Cn & 10.45   & 4 &  -0.89   &  -1.45    &   -1.19   &  -1.31  &  -1.03    &  -0.98   &  -2.64   &  SF        &  $\alpha$   \\
$^{277}$Ds & 10.83   & 4 &  -2.39   &  -3.22    &   -3.01   &  -2.83  &  -2.68    &  -1.99   &  1.90    &  $\alpha$  &  $\alpha$   \\
$^{273}$Hs & 9.70    & 5 &  -0.12   &  -0.59    &   -0.40   &  -0.45  &  -0.34    &  0.11    &  4.68    &  $\alpha$  &  $\alpha$   \\
$^{269}$Sg & 8.65    & 5 &  2.27    &  2.21     &   2.36    &  2.12   &  1.88     &  2.06    &  5.35    &  $\alpha$  &  $\alpha$   \\
\hline
$^{294}$Og & 11.60   & 0 &  -3.05   &  -3.80    &   -2.82   &  -1.72  &  -0.40    &  -1.59   &  6.15    &  $\alpha$  &  $\alpha$    \\
$^{290}$Lv & 11.00   & 0  &  -1.82   &  -2.67    &   -1.81   &  -0.94  &  -1.20    &  -1.44   &  4.41    &  $\alpha$  &  $\alpha$    \\
$^{286}$Fl & 10.37   & 0  &  -0.80   &  -1.36    &   -0.64   &  -0.03  &  -0.18    &  -0.96   &  1.16    &  $\alpha$  &  $\alpha$/SF \\
$^{282}$Cn & 10.17   & 0  &  -3.30   &  -1.48    &   -0.78   &  -0.01  &  -0.31    &  -0.51   &  -4.26   &  SF        &  SF          \\
$^{278}$Ds & 10.47   & 0  &          &  -3.24    &   -2.44   &  -1.05  &  -1.78    &  -1.20   &  -4.39   &  SF        &              \\
$^{274}$Hs & 9.57    & 0  &          &  -1.04    &   -0.43   &  0.51   &  -0.01    &  -0.01   &  0.45    &  $\alpha$  &              \\
$^{270}$Sg & 9.00    & 0  &          &  0.16     &   0.67    &  1.44   &  0.83     &  1.65    &  2.74    &  $\alpha$  &              \\
\hline                                                                                                                          \hline
\end{tabular}}
\label{prediction1}
\end{table}

\begin{table}[htbp]
\caption{Continue to Table \ref{prediction1}.}
\centering
\resizebox{0.9\textwidth}{!}{%
\begin{tabular}{c|ccc|ccccc|c|cc}
\hline
\hline
\multicolumn{1}{c|}{Nucleus}&
\multicolumn{3}{c|}{Expt.}&
\multicolumn{5}{c|}{log$_{10}$T$_{1/2}^{\alpha}$(s)}&
\multicolumn{1}{c|}{log$_{10}$T$_{1/2}^{SF}$(s)}&
\multicolumn{2}{c}{Decay Modes}\\
\cline{2-11}
\multicolumn{1}{c|}{}&
\multicolumn{1}{c}{Q$_{\alpha}$}&
\multicolumn{1}{c}{l}&
\multicolumn{1}{c|}{log$_{10}$T$_{1/2}$(s)}&
\multicolumn{1}{c}{MHF (Set-1)}&
\multicolumn{1}{c}{MHF (Set-2)}&
\multicolumn{1}{c}{MSF}&
\multicolumn{1}{c}{XGBoost}&
\multicolumn{1}{c|}{MLP}&
\multicolumn{1}{c|}{MBF}&
\multicolumn{1}{c}{Predicted}&
\multicolumn{1}{c}{Expt.}\\
\hline
$^{295}$Og &  10.88*  & 0 &         &  -0.65  &  -0.32  &  -0.72  &  -0.24   & -1.62   &  8.45   &  $\alpha$   &                  \\
$^{291}$Lv &  10.89   & 2 &  -2.20  &  -1.38  &  -1.10  &  -1.30  &  -0.94   & -1.69   &  7.59   &  $\alpha$   &  $\alpha$        \\
$^{287}$Fl &  10.16   & 0 &  -0.29  &  0.12   &  0.41   &  0.07   &  0.36    & -0.04   &  4.64   &  $\alpha$   &  $\alpha$        \\
$^{283}$Cn &  9.94    & 7 &  0.60   &  0.08   &  0.34   &  0.10   &  0.29    & 0.32    &  -0.56  &  SF         &  $\alpha$/SF     \\
$^{279}$Ds &  10.08   & 7 &  -0.74  &  -1.10  &  -0.89  &  -0.88  &  -0.76   & -0.22   &  -3.10  &  SF         &  SF              \\
$^{275}$Hs &  9.44    & 0 &  -0.82  &  0.20   &  0.38   &  0.32   &  0.30    & 0.13    &  1.54   &  $\alpha$   &  $\alpha$        \\
$^{271}$Sg &  8.89    & 5 &  2.16   &  1.31   &  1.44   &  1.37   &  1.08    & 2.00    &  3.74   &  $\alpha$   &  $\alpha$        \\
\hline
$^{296}$Og &  10.71*  & 0 &         &  -1.02  &  -0.23  &  0.22   &  0.19    & -1.36   &  5.65   &  $\alpha$   &                  \\
$^{292}$Lv &  10.77   & 0  &  -1.74  &  -2.04  &  -1.20  &  -0.31  &  -0.65   & -1.46   &  4.81   &  $\alpha$   &  $\alpha$        \\
$^{288}$Fl &  10.07   & 0  &  -0.28  &  -0.43  &  0.25   &  0.80   &  0.56    & 0.02    &  2.42   &  $\alpha$   &  $\alpha$        \\
$^{284}$Cn &  9.60    & 0 &  -1.00  &  0.44   &  1.03   &  1.48   &  1.15    & 0.73    &  -2.49  &  SF         &  SF              \\
$^{280}$Ds &  9.81    & 0  &         &  -1.18  &  -0.49  &  0.58   &  -0.04   & 0.25    &  -5.25  &  SF         &                  \\
$^{276}$Hs &  9.28    & 0  &         &  -0.13  &  0.47   &  1.40   &  0.63    & 0.41    &  -4.96  &  SF         &                  \\
$^{272}$Sg &  8.70    & 0  &         &  1.18   &  1.68   &  2.43   &  1.69    & 1.13    &  -1.75  &  SF         &                  \\
\hline
$^{297}$Og &  10.52*  & 0 &         &  0.41   &  0.76   &  0.25   &  0.66    & -1.12   &  8.31   &  $\alpha$   &                  \\
$^{293}$Lv &  10.68   & 2 &  -1.28  &  -0.81  &  -0.53  &  -0.73  &  -0.43   & -1.19   &  8.10   &  $\alpha$   &  $\alpha$        \\
$^{289}$Fl &  9.97    & 0 &  -0.01  &  0.70   &  0.98   &  0.63   &  0.80    & 0.05    &  6.00   &  $\alpha$   &  $\alpha$        \\
$^{285}$Cn &  9.32    & 2 &  1.48   &  2.14   &  2.41   &  1.96   &  1.91    & 1.06    &  1.55   &  $\alpha$   &  $\alpha$        \\
$^{281}$Ds &  9.51    & 0 &  1.11   &  0.69   &  0.90   &  0.76   &  0.67    & 0.70    &  -1.83  &  SF         &  SF/$\alpha$     \\
$^{277}$Hs &  9.05    & 0 &  -2.52  &  1.50   &  1.66   &  1.54   &  1.24    & -0.71   &  -2.69  &  SF         &  $\alpha$        \\
$^{273}$Sg &  8.20*   & 0 &         &  3.89   &  4.00   &  3.73   &  3.52    & 2.32    &  -0.51  &  SF         &                  \\
\hline
$^{298}$Og &  10.40*  & 0 &         &  -0.04  &  0.70   &  1.06   &  0.93    & -1.11   &  5.18   &  $\alpha$   &                  \\
$^{294}$Lv &  9.90*   & 0 &         &  0.98   &  1.61   &  1.81   &  1.61    & 0.07    &  5.75   &  $\alpha$   &                  \\
$^{290}$Fl &  9.39*   & 0 &         &  2.07   &  2.60   &  2.63   &  2.34    & 0.68    &  4.16   &  $\alpha$   &                  \\
$^{286}$Cn &  9.07*   & 0 &         &  2.41   &  2.90   &  3.01   &  2.55    & 1.11    &  -0.60  &  SF         &                  \\
$^{282}$Ds &  9.04*   & 0 &         &  1.57   &  2.11   &  2.64   &  1.98    & -0.09   &  -4.17  &  SF         &                  \\
$^{278}$Hs &  8.47*   & 0 &         &  2.96   &  3.40   &  3.72   &  3.28    & 0.57    &  -7.12  &  SF         &                  \\
$^{274}$Sg &  8.13*   & 0 &         &  3.42   &  3.83   &  4.23   &  3.82    & 2.32    &  -6.37  &  SF         &                  \\
\hline
$^{299}$Og &  10.02*  & 2 &         &  2.04   &  2.42   &  1.71   &  1.89    & 0.05    &  7.57   &  $\alpha$   &                  \\
$^{295}$Lv &  9.54*   & 0 &         &  2.91   &  3.27   &  2.53   &  2.52    & 0.59    &  8.86   &  $\alpha$   &                  \\
$^{291}$Fl &  9.23*   & 0 &         &  3.24   &  3.57   &  2.90   &  2.75    & 0.86    &  7.52   &  $\alpha$   &                  \\
$^{287}$Cn &  9.03*   & 2 &         &  3.15   &  3.42   &  2.90   &  2.68    & 0.48    &  3.39   &  $\alpha$   &                  \\
$^{283}$Ds &  8.84*   & 2 &         &  3.04   &  3.25   &  2.88   &  2.65    & 0.54    &  -0.62  &  SF         &                  \\
$^{279}$Hs &  8.48*   & 0 &         &  3.60   &  3.76   &  3.45   &  3.30    & 0.67    &  -3.67  &  SF         &                  \\
$^{275}$Sg &  8.07*   & 4 &         &  4.40   &  4.50   &  4.25   &  4.11    & 2.33    &  -3.33  &  SF         &                  \\
\hline
$^{300}$Og &  9.75*   & 0 &         &  2.36   &  2.94   &  2.77   &  2.58    & 0.34    &  3.98   &  $\alpha$   &                  \\
$^{296}$Lv &  9.33*   & 0 &         &  3.14   &  3.64   &  3.40   &  3.11    & 0.80    &  5.74   &  $\alpha$   &                  \\
$^{292}$Fl &  9.09*   & 0 &         &  3.18   &  3.67   &  3.59   &  3.23    & 0.49    &  5.71   &  $\alpha$   &                  \\
$^{288}$Cn &  8.83*   & 0 &         &  3.30   &  3.77   &  3.85   &  3.45    & 0.51    &  -0.16  &  SF         &                  \\
$^{284}$Ds &  8.86*   & 0 &         &  2.20   &  2.76   &  3.35   &  2.64    & 0.54    &  -2.83  &  SF         &                  \\
$^{280}$Hs &  8.37*   & 0 &         &  3.29   &  3.78   &  4.26   &  3.77    & 1.09    &  -6.01  &  SF         &                  \\
$^{276}$Sg &  7.98*   & 0 &         &  3.98   &  4.42   &  4.93   &  4.50    & 2.37    &  -7.43  &  SF         &                  \\
\hline
$^{301}$Og &  9.47*   & 2 &         &  3.95   &  4.38   &  3.40   &  3.36    & 0.62    &  6.06   &  $\alpha$   &                  \\
$^{297}$Lv &  9.19*   & 2 &         &  4.19   &  4.58   &  3.69   &  3.62    & 0.62    &  8.49   &  $\alpha$   &                  \\
$^{293}$Fl &  8.92*   & 0 &         &  4.41   &  4.75   &  3.96   &  3.89    & 0.16    &  8.68   &  $\alpha$   &                  \\
$^{289}$Cn &  8.85*   & 0 &         &  3.82   &  4.09   &  3.54   &  3.44    & 0.63    &  2.21   &  $\alpha$   &                  \\
$^{285}$Ds &  8.80*   & 2 &         &  3.16   &  3.36   &  3.05   &  2.91    & 0.49    &  1.34   &  $\alpha$   &                  \\
$^{281}$Hs &  8.33*   & 2 &         &  4.17   &  4.32   &  4.02   &  3.96    & 1.17    &  -1.79  &  SF         &                  \\
$^{277}$Sg &  7.92*   & 0 &         &  4.98   &  5.07   &  4.83   &  4.80    & 2.40    &  -5.15  &  SF         &                  \\
\hline
$^{302}$Og &  9.34*   & 0 &         &  3.98   &  4.46   &  4.00   &  3.82    & 0.70    &  1.96   &  $\alpha$   &                  \\
$^{298}$Lv &  9.00*   & 0 &         &  4.48   &  4.91   &  4.48   &  4.33    & -0.06   &  5.01   &  $\alpha$   &                  \\
$^{294}$Fl &  8.91*   & 0 &         &  3.87   &  4.36   &  4.30   &  3.98    & 0.37    &  5.69   &  $\alpha$   &                  \\
$^{290}$Cn &  8.75*   & 0 &         &  3.55   &  4.07   &  4.31   &  3.87    & 0.64    &  0.09   &  SF         &                  \\
$^{286}$Ds &  8.77*   & 0 &         &  2.41   &  3.03   &  3.81   &  3.07    & 0.49    &  -1.44  &  SF         &                  \\
$^{282}$Hs &  8.25*   & 0 &         &  3.67   &  4.19   &  4.83   &  4.27    & 1.19    &  -4.08  &  SF         &                  \\
$^{278}$Sg &  7.74*   & 0 &         &  4.98   &  5.43   &  5.93   &  5.71    & 2.88    &  -7.64  &  SF         &                  \\
\hline
\hline
\end{tabular}}
\label{prediction2}
\end{table}

\begin{table}[htbp]
\caption{Prediction of decay chains from modified formulas and machine learning for Z$=$119 isotopes. Q$_{\alpha}$  and log$_{10}$T$_{1/2}$ are taken from Ref. \cite{nndc}, whereas (*) values of Q$_{\alpha}$ are taken from XGBoost algorithm as described in text. The angular momentum ($l$) is calculated from selection rule \cite{denisov2009,royer2020} based on parity and spin of parent and daughter nuclei, which are taken from latest evaluated nuclear properties table NUBASE2016 \cite{audi2017} or P. M\"{o}ller \cite{mollerparity}.}
\centering
\resizebox{0.85\textwidth}{!}{%
\begin{tabular}{c|ccc|ccccc|c|cc}
\hline
\hline
\multicolumn{1}{c|}{Nucleus}&
\multicolumn{3}{c|}{Expt.}&
\multicolumn{5}{c|}{log$_{10}$T$_{1/2}^{\alpha}$(s)}&
\multicolumn{1}{c|}{log$_{10}$T$_{1/2}^{SF}$(s)}&
\multicolumn{2}{c}{Decay Modes}\\
\cline{2-11}
\multicolumn{1}{c|}{}&
\multicolumn{1}{c}{Q$_{\alpha}$}&
\multicolumn{1}{c}{l}&
\multicolumn{1}{c|}{log$_{10}$T$_{1/2}$(s)}&
\multicolumn{1}{c}{MHF (Set-1)}&
\multicolumn{1}{c}{MHF (Set-2)}&
\multicolumn{1}{c}{MSF}&
\multicolumn{1}{c}{XGBoost}&
\multicolumn{1}{c|}{MLP}&
\multicolumn{1}{c|}{MBF}&
\multicolumn{1}{c}{Predicted}&
\multicolumn{1}{c}{Expt.}\\
\hline
$^{287}$119 & 11.21*   & 2 &          &  -0.73  &  -0.82   &  -0.83   &  -0.92  & -2.70  &  6.02     &   $\alpha$   &               \\
$^{283}$Ts  & 11.29*   & 2 &          &  -1.49  &  -1.73   &  -1.64   &  -1.75  & -3.12  &  4.00     &   $\alpha$   &               \\
$^{279}$Mc  & 11.39*   & 4 &          &  -2.30  &  -2.68   &  -2.50   &  -2.62  & -3.37  &  1.64     &   $\alpha$   &               \\
$^{275}$Nh  & 11.05*   & - &          &  -2.13  &  -2.41   &  -2.28   &  -2.28  & -2.78  &  4.87     &   $\alpha$   &               \\
$^{271}$Rg  & 10.95*   & 1 &          &  -2.49  &  -2.79   &  -2.64   &  -2.60  & -2.78  &  3.36     &   $\alpha$   &               \\
$^{267}$Mt  & 10.90    & 3 &          &  -2.99  &  -3.31   &  -3.15   &  -3.08  & -3.02  &  3.32     &   $\alpha$   &               \\
$^{263}$Bh  & 10.10    & 2 &          &  -1.71  &  -1.80   &  -1.76   &  -1.56  & -1.14  &  1.42     &   $\alpha$   &               \\
\hline
$^{288}$119 & 11.92*   & -  &          &  -1.71  &  -2.42   &  -1.80   &  -0.82  & -2.71  &  7.49     &   $\alpha$   &               \\
$^{284}$Ts  & 11.88*   & -  &          &  -2.15  &  -2.96   &  -2.21   &  -1.23  & -3.21  &  5.75     &   $\alpha$   &               \\
$^{280}$Mc  & 11.63*   & -  &          &  -2.19  &  -2.95   &  -2.23   &  -1.54  & -3.44  &  2.98     &   $\alpha$   &               \\
$^{276}$Nh  & 11.40*   & -  &          &  -2.27  &  -2.99   &  -2.29   &  -1.65  & -3.51  &  7.67     &   $\alpha$   &               \\
$^{272}$Rg  & 11.20    & -  & -2.42    &  -2.40  &  -3.11   &  -2.40   &  -1.76  & -3.58  &  6.42     &   $\alpha$   &   $\alpha$    \\
$^{268}$Mt  & 10.67    & -  & -1.68    &  -1.86  &  -2.36   &  -1.86   &  -1.63  & -1.92  &  5.22     &   $\alpha$   &   $\alpha$    \\
$^{264}$Bh  & 9.96     & -  & -0.36    &  -0.87  &  -1.05   &  -0.87   &  -0.56  & -0.92  &  3.55     &   $\alpha$   &   $\alpha$    \\
\hline
$^{289}$119 & 11.19*   & 2 &          &  -0.72  &  -0.80   &  -0.81   &  -0.76  & -2.74  &  7.65     &   $\alpha$   &               \\
$^{285}$Ts  & 11.17*   & 2 &          &  -1.24  &  -1.42   &  -1.36   &  -1.30  & -2.73  &  5.98     &   $\alpha$   &               \\
$^{281}$Mc  & 11.14*   & 2 &          &  -1.77  &  -2.05   &  -1.91   &  -1.85  & -2.86  &  3.08     &   $\alpha$   &               \\
$^{277}$Nh  & 11.36*   & 0 &          &  -2.83  &  -3.30   &  -3.04   &  -3.04  & -3.19  &  5.32     &   $\alpha$   &               \\
$^{273}$Rg  & 10.90    & 5 &          &  -2.40  &  -2.72   &  -2.54   &  -2.42  & -2.65  &  6.90     &   $\alpha$   &               \\
$^{269}$Mt  & 10.50    & 3 &          &  -2.07  &  -2.30   &  -2.16   &  -2.01  & -2.06  &  5.62     &   $\alpha$   &               \\
$^{265}$Bh  & 9.68     & 2 & -0.05    &  -0.63  &  -0.62   &  -0.61   &  -0.46  & -0.14  &  3.34     &   $\alpha$   &   $\alpha$    \\
\hline
$^{290}$119 & 11.16*   & -  &          &  -0.19  &  -0.35   &  -0.29   &  -0.62  & -2.64  &  9.45     &   $\alpha$   &               \\
$^{286}$Ts  & 11.12*   & -  &          &  -0.65  &  -0.95   &  -0.73   &  -1.14  & -2.66  &  7.52     &   $\alpha$   &               \\
$^{282}$Mc  & 11.00*   & -  &          &  -0.93  &  -1.29   &  -0.98   &  -1.42  & -2.73  &  4.71     &   $\alpha$   &               \\
$^{278}$Nh  & 11.85    & -  & -3.62    &  -3.15  &  -4.20   &  -3.14   &  -4.37  & -3.64  &  6.88     &   $\alpha$   &   $\alpha$    \\
$^{274}$Rg  & 11.48    & -  & -2.19    &  -2.97  &  -3.90   &  -2.94   &  -3.97  & -2.85  &  9.62     &   $\alpha$   &   $\alpha$    \\
$^{270}$Mt  & 10.18    & -  & -2.30    &  -0.80  &  -1.05   &  -0.81   &  -1.17  & -1.54  &  7.08     &   $\alpha$   &   $\alpha$    \\
$^{266}$Bh  & 9.43     & -  & 0.23     &  0.40   &  0.50    &  0.38    &  0.19   & 0.49   &  4.94     &   $\alpha$   &   $\alpha$    \\
\hline
$^{291}$119 & 11.10*   & 2 &          &  -0.53  &  -0.57   &  -0.60   &  -0.43  & -2.59  &  9.72     &   $\alpha$   &               \\
$^{287}$Ts  & 11.11*   & 2 &          &  -1.12  &  -1.30   &  -1.23   &  -1.07  & -2.65  &  7.35     &   $\alpha$   &               \\
$^{283}$Mc  & 10.89*   & 2 &          &  -1.19  &  -1.37   &  -1.28   &  -1.11  & -2.47  &  4.93     &   $\alpha$   &               \\
$^{279}$Nh  & 11.50    & 0 &          &  -3.13  &  -3.73   &  -3.37   &  -3.36  & -2.94  &  -0.81    &   $\alpha$   &               \\
$^{275}$Rg  & 11.80    & 5 &          &  -4.33  &  -5.11   &  -4.62   &  -4.93  & -3.30  &  8.23     &   $\alpha$   &               \\
$^{271}$Mt  & 9.91     & 3 &          &  -0.60  &  -0.64   &  -0.59   &  -0.47  & -0.47  &  8.81     &   $\alpha$   &               \\
$^{267}$Bh  & 9.23     & 2 & 1.23     &  0.62   &  0.76    &  0.71    &  0.69   & 0.98   &  6.00     &   $\alpha$   &  $\alpha$     \\
\hline
$^{292}$119 & 11.11*   & -  &          &  -0.11  &  -0.24   &  -0.19   &  -0.41  & -2.62  &  11.28    &   $\alpha$   &               \\
$^{288}$Ts  & 11.05*   & -  &          &  -0.53  &  -0.80   &  -0.59   &  -0.92  & -2.33  &  8.87     &   $\alpha$   &               \\
$^{284}$Mc  & 10.74*   & -  &          &  -0.40  &  -0.61   &  -0.45   &  -0.75  & -1.90  &  6.16     &   $\alpha$   &               \\
$^{280}$Nh  & 11.20    & -  &          &  -1.90  &  -2.60   &  -1.89   &  -1.45  & -2.01  &  0.32     &   $\alpha$   &               \\
$^{276}$Rg  & 11.50    & -  &          &  -2.12  &  -2.87   &  -2.10   &  -2.85  & -1.90  &  6.42     &   $\alpha$   &               \\
$^{272}$Mt  & 10.05    & -  &          &  -0.50  &  -0.72   &  -0.50   &  -0.87  & -0.34  &  9.74     &   $\alpha$   &               \\
$^{268}$Bh  & 9.00     & -  &          &  1.52   &  1.88    &  1.50    &  1.27   & 1.07   &  7.56     &   $\alpha$   &               \\
\hline
$^{293}$119 & 11.09*   & 2 &          &  -0.51  &  -0.57   &  -0.57   &  -0.36  & -2.49  &  11.75    &   $\alpha$   &               \\
$^{289}$Ts  & 10.99*   & 2 &          &  -0.85  &  -1.01   &  -0.94   &  -0.76  & -2.06  &  8.50     &   $\alpha$   &               \\
$^{285}$Mc  & 10.63*   & 4 &          &  -0.58  &  -0.66   &  -0.62   &  -0.46  & -1.58  &  6.02     &   $\alpha$   &               \\
$^{281}$Nh  & 11.00    & 2 &          &  -2.03  &  -2.45   &  -2.18   &  -2.10  & -1.70  &  0.44     &   $\alpha$   &               \\
$^{277}$Rg  & 11.20    & 1 &          &  -3.06  &  -3.68   &  -3.27   &  -3.26  & -1.86  &  4.56     &   $\alpha$   &               \\
$^{273}$Mt  & 10.81    & 3 &          &  -2.78  &  -3.30   &  -2.94   &  -2.90  & -1.83  &  8.08     &   $\alpha$   &               \\
$^{269}$Bh  & 8.60     & 2 &          &  2.53   &  2.93    &  2.74    &  2.31   & 1.58   &  7.61     &   $\alpha$   &               \\
\hline
$^{294}$119 & 11.04*   & -  &          &  0.03   &  -0.08   &  -0.04   &  -0.26  & -2.22  &  13.24    &   $\alpha$   &               \\
$^{290}$Ts  & 10.98*   & -  &          &  -0.37  &  -0.63   &  -0.42   &  -0.76  & -1.96  &  10.03    &   $\alpha$   &               \\
$^{286}$Mc  & 10.58*   & -  &          &  -0.03  &  -0.16   &  -0.07   &  -0.35  & -1.14  &  7.36     &   $\alpha$   &               \\
$^{282}$Nh  & 10.75    & -  & -1.15    &  -0.96  &  -1.41   &  -0.96   &  -1.48  & -1.20  &  2.13     &   $\alpha$   &   $\alpha$    \\
$^{278}$Rg  & 10.47    & -  & -2.38    &  -0.88  &  -1.31   &  -0.87   &  -1.39  & -1.25  &  1.72     &   $\alpha$   &   $\alpha$     \\
$^{274}$Mt  & 10.60    & -  & -0.36    &  -1.72  &  -2.40   &  -1.66   &  -2.38  & -1.27  &  8.86     &   $\alpha$   &   $\alpha$    \\
$^{270}$Bh  & 9.06     & -  & 1.78     &  1.39   &  1.63    &  1.38    &  1.09   & 1.20   &  9.45     &   $\alpha$   &   $\alpha$    \\
\hline
$^{295}$119 & 11.01*   & 2 &          &  -0.33  &  -0.39   &  -0.38   &  -0.21  & -2.12  &  12.05    &   $\alpha$   &               \\
$^{291}$Ts  & 11.50    & 2 &          &  -1.99  &  -2.53   &  -2.20   &  -2.14  & -2.16  &  10.10    &   $\alpha$   &               \\
$^{287}$Mc  & 10.76    & 2 & -1.49    &  -0.89  &  -1.11   &  -0.96   &  -0.87  & -1.23  &  6.97     &   $\alpha$   &   $\alpha$    \\
$^{283}$Nh  & 10.51    & 2 & -1.00    &  -0.88  &  -1.10   &  -0.93   &  -0.86  & -0.92  &  2.13     &   $\alpha$   &   $\alpha$    \\
$^{279}$Rg  & 10.52    & 1 & -0.77    &  -1.49  &  -1.86   &  -1.58   &  -1.55  & -0.95  &  0.36     &   $\alpha$   &   $\alpha$    \\
$^{275}$Mt  & 10.48    & 5 & -2.01    &  -1.99  &  -2.47   &  -2.10   &  -2.10  & -1.40  &  5.14     &   $\alpha$   &   $\alpha$    \\
$^{271}$Bh  & 9.42     & 0 &          &  0.12   &  0.02    &  0.16    &  0.09   & 0.88   &  7.60     &   $\alpha$   &               \\
\hline
\hline
\end{tabular}}
\label{prediction3}
\end{table}

\begin{table}[htbp]
\caption{Continue to Table \ref{prediction3}.}
\centering
\resizebox{0.9\textwidth}{!}{%
\begin{tabular}{c|ccc|ccccc|c|cc}
\hline
\hline
\multicolumn{1}{c|}{Nucleus}&
\multicolumn{3}{c|}{Expt.}&
\multicolumn{5}{c|}{log$_{10}$T$_{1/2}^{\alpha}$(s)}&
\multicolumn{1}{c|}{log$_{10}$T$_{1/2}^{SF}$(s)}&
\multicolumn{2}{c}{Decay Modes}\\
\cline{2-11}
\multicolumn{1}{c|}{}&
\multicolumn{1}{c}{Q$_{\alpha}$}&
\multicolumn{1}{c}{l}&
\multicolumn{1}{c|}{log$_{10}$T$_{1/2}$(s)}&
\multicolumn{1}{c}{MHF (Set-1)}&
\multicolumn{1}{c}{MHF (Set-2)}&
\multicolumn{1}{c}{MSF}&
\multicolumn{1}{c}{XGBoost}&
\multicolumn{1}{c|}{MLP}&
\multicolumn{1}{c|}{MBF}&
\multicolumn{1}{c}{Predicted}&
\multicolumn{1}{c}{Expt.}\\
\hline
$^{296}$119 & 11.03*  & -  &         &  0.05    &  -0.11   &  -0.01   &  -0.30   &  -1.67   &   11.68   &  $\alpha$   &              \\
$^{292}$Ts  & 10.14   & -  &         &  1.56    &  1.97    &  1.48    &  1.33    &  -0.19   &   11.21   &  $\alpha$   &              \\
$^{288}$Mc  & 10.75   & -  &  -1.06  &  -0.41   &  -0.74   &  -0.42   &  -0.87   &  -1.18   &   8.31    &  $\alpha$   &  $\alpha$    \\
$^{284}$Nh  & 10.28   & -  &  -0.32  &  0.11    &  -0.05   &  0.10    &  -0.27   &  -0.53   &   3.82    &  $\alpha$   &  $\alpha$    \\
$^{280}$Rg  & 11.20   & -  &  -2.38  &  -2.41   &  -3.44   &  -2.33   &  -0.37   &  -0.62   &   -0.61   &  $\alpha$   &  $\alpha$    \\
$^{276}$Mt  & 10.10   & -  &  -0.14  &  -0.59   &  -1.02   &  -0.54   &  -1.12   &  -0.39   &   4.29    &  $\alpha$   &  $\alpha$    \\
$^{272}$Bh  & 9.30    & -  &  1.00   &  0.79    &  0.75    &  0.81    &  0.36    &  1.08    &   6.69    &  $\alpha$   &  $\alpha$    \\
\hline
$^{297}$119 & 10.93*  & 2 &         &  -0.14   &  -0.19   &  -0.17   &  -0.05   &  -1.61   &   9.97    &  $\alpha$   &              \\
$^{293}$Ts  & 11.29   & 2 &  -1.85  &  -1.53   &  -2.01   &  -1.69   &  -1.65   &  -1.87   &   9.45    &  $\alpha$   &  $\alpha$    \\
$^{289}$Mc  & 10.51   & 4 &  -0.66  &  -0.28   &  -0.41   &  -0.29   &  -0.25   &  -0.91   &   7.91    &  $\alpha$   &  $\alpha$    \\
$^{285}$Nh  & 10.01   & 3 &  0.74   &  0.39    &  0.41    &  0.45    &  0.43    &  0.28    &   3.52    &  $\alpha$   &  $\alpha$    \\
$^{281}$Rg  & 9.90    & 7 &  1.23   &  0.08    &  -0.02   &  0.12    &  0.07    &  0.22    &   -0.17   &  SF         &  SF          \\
$^{277}$Mt  & 9.91    & 0 &  0.70   &  -0.55   &  -0.84   &  -0.56   &  -0.63   &  -0.13   &   -0.33   &  $\alpha$   &  SF          \\
$^{273}$Bh  & 9.10    & 2 &         &  1.05    &  1.03    &  1.15    &  0.85    &  1.09    &   4.31    &  $\alpha$   &              \\
\hline
$^{298}$119 & 10.88*  & -  &         &  0.39    &  0.31    &  0.34    &  0.06    &  -1.64   &   10.64   &  $\alpha$   &              \\
$^{294}$Ts  & 10.81   & -  &  -1.10  &  0.01    &  1.02    &  -0.01    &  0.68    &  -1.08   &   10.82   &  $\alpha$   &  $\alpha$    \\
$^{290}$Mc  & 10.40   & -  &  -1.80  &  0.40    &  0.28    &  0.38    &  0.02    &  -1.15   &   9.19    &  $\alpha$   &  $\alpha$    \\
$^{286}$Nh  & 9.79    & -  &  1.30   &  1.31    &  1.49    &  1.29    &  0.96    &  0.59    &   5.37    &  $\alpha$   &  $\alpha$     \\
$^{282}$Rg  & 9.64    & -  &  -0.30  &  0.81    &  0.47    &  0.77    &  0.71    &  0.44    &   0.72    &  $\alpha$   &  $\alpha$     \\
$^{278}$Mt  & 9.63    & -  &  0.90   &  0.57    &  0.40    &  0.60    &  0.12    &  0.34    &   -0.81   &  SF         &  $\alpha$    \\
$^{274}$Bh  & 8.95    & -  &  -0.05  &  1.75    &  1.90    &  1.77    &  1.22    &  0.98    &   2.24    &  $\alpha$   &  $\alpha$    \\
\hline
$^{299}$119 & 10.73*  & 2 &         &  0.33    &  0.37    &  0.35    &  0.43    &  -1.30   &   9.44    &  $\alpha$   &              \\
$^{295}$Ts  & 10.48*  & 2 &         &  0.39    &  0.40    &  0.42    &  0.44    &  -1.15   &   10.13   &  $\alpha$   &              \\
$^{291}$Mc  & 10.30   & 2 &         &  0.26    &  0.20    &  0.29    &  0.27    &  -0.83   &   8.85    &  $\alpha$   &              \\
$^{287}$Nh  & 9.54    & 3 &         &  1.68    &  1.96    &  1.85    &  1.63    &  0.75    &   5.07    &  $\alpha$   &              \\
$^{283}$Rg  & 9.36    & 2 &         &  1.58    &  1.75    &  1.73    &  1.47    &  0.91    &   0.61    &  SF         &              \\
$^{279}$Mt  & 9.40    & 0 &         &  0.85    &  0.76    &  0.94    &  0.59    &  0.46    &   -2.07   &  SF         &              \\
$^{275}$Bh  & 8.80*   & 3 &         &  1.99    &  2.04    &  2.14    &  1.72    &  0.67    &   -0.93   &  SF         &              \\
\hline
$^{300}$119 & 10.47*  & -  &         &  1.35    &  1.57    &  1.29    &  1.07    &  -1.15   &   9.90    &  $\alpha$   &              \\
$^{296}$Ts  & 10.25*  & -  &         &  1.33    &  1.50    &  1.29    &  1.01    &  -0.73   &   11.10   &  $\alpha$   &              \\
$^{292}$Mc  & 9.64*   & -  &         &  2.29    &  2.79    &  2.25    &  1.97    &  0.33    &   10.30   &  $\alpha$   &              \\
$^{288}$Nh  & 9.21*   & -  &         &  2.85    &  3.50    &  2.81    &  2.50    &  0.94    &   6.39    &  $\alpha$   &              \\
$^{284}$Rg  & 8.94*   & -  &         &  3.03    &  3.65    &  3.00    &  2.64    &  1.06    &   2.25    &  $\alpha$   &              \\
$^{280}$Mt  & 8.83*   & -  &         &  2.72    &  3.15    &  2.71    &  2.30    &  0.27    &   -0.56   &  SF         &              \\
$^{276}$Bh  & 8.42*   & -  &         &  3.33    &  3.85    &  3.32    &  3.12    &  1.67    &   -0.79   &  SF         &              \\
\hline
$^{301}$119 & 10.25*  & 0 &         &  1.56    &  1.92    &  1.72    &  1.60    &  -0.73   &   8.12    &  $\alpha$   &              \\
$^{297}$Ts  & 9.92*   & 2 &         &  1.85    &  2.24    &  2.04    &  1.83    &  0.06    &   10.02   &  $\alpha$   &              \\
$^{293}$Mc  & 9.46*   & 2 &         &  2.55    &  3.07    &  2.80    &  2.45    &  0.62    &   10.18   &  $\alpha$   &              \\
$^{289}$Nh  & 9.11*   & 7 &         &  2.95    &  3.47    &  3.23    &  2.76    &  0.91    &   5.63    &  $\alpha$   &              \\
$^{285}$Rg  & 8.82*   & 2 &         &  3.21    &  3.69    &  3.50    &  3.08    &  0.36    &   2.79    &  $\alpha$   &              \\
$^{281}$Mt  & 8.83*   & 0 &         &  2.57    &  2.76    &  2.77    &  2.35    &  0.55    &   -0.12   &  SF         &              \\
$^{277}$Bh  & 8.27*   & 3 &         &  3.73    &  4.04    &  3.99    &  3.72    &  1.81    &   -2.78   &  SF         &              \\
\hline
$^{302}$119 & 9.92*   & -  &         &  2.73    &  3.43    &  2.65    &  2.43    &  0.06    &   8.20    &  $\alpha$   &              \\
$^{298}$Ts  & 9.70*   & -  &         &  2.74    &  3.39    &  2.68    &  2.41    &  0.35    &   10.56   &  $\alpha$   &              \\
$^{294}$Mc  & 9.27*   & -  &         &  3.31    &  4.10    &  3.26    &  2.93    &  0.83    &   11.39   &  $\alpha$   &              \\
$^{290}$Nh  & 8.93*   & -  &         &  3.69    &  4.54    &  3.64    &  3.45    &  0.07    &   7.21    &  $\alpha$   &              \\
$^{286}$Rg  & 8.87*   & -  &         &  3.26    &  3.85    &  3.24    &  2.94    &  0.56    &   4.18    &  $\alpha$   &              \\
$^{282}$Mt  & 8.72*   & -  &         &  3.09    &  3.52    &  3.09    &  2.79    &  0.58    &   1.30    &  $\alpha$   &              \\
$^{278}$Bh  & 8.21*   & -  &         &  4.02    &  4.64    &  4.01    &  4.00    &  1.85    &   -2.40   &  SF         &              \\
\hline
$^{303}$119 & 9.68*   & 2 &         &  3.11    &  3.92    &  3.44    &  3.06    &  0.35    &   5.99    &  $\alpha$   &              \\
$^{299}$Ts  & 9.44*   & 2 &         &  3.23    &  3.98    &  3.56    &  3.12    &  0.65    &   9.02    &  $\alpha$   &              \\
$^{295}$Mc  & 9.15*   & 2 &         &  3.49    &  4.21    &  3.83    &  3.42    &  -0.24   &   10.31   &  $\alpha$   &              \\
$^{291}$Nh  & 8.85*   & 7 &         &  3.80    &  4.48    &  4.15    &  3.81    &  0.58    &   6.75    &  $\alpha$   &              \\
$^{287}$Rg  & 8.79*   & 3 &         &  3.37    &  3.79    &  3.65    &  3.32    &  0.59    &   3.07    &  $\alpha$   &              \\
$^{283}$Mt  & 8.70*   & 2 &         &  3.03    &  3.22    &  3.25    &  2.94    &  0.59    &   0.99    &  $\alpha$   &              \\
$^{279}$Bh  & 8.21*   & 0 &         &  3.96    &  4.22    &  4.23    &  4.01    &  1.87    &   -2.84   &  SF         &              \\
\hline
\hline
\end{tabular}}
\label{prediction4}
\end{table}

\section{Conclusions}
Empirical formulas for $\alpha$-decay half-life (Horoi scaling law and Sobiczewski formula) and spontaneous fission half-life (Bao \textit{et al.}) are modified and the coefficients are fitted using latest evaluated half-lives in the range 82$\leq$Z$\leq$118. These formulas are found with a good accuracy while compared with the original formulas and various other formulas, and therefore, utilized to demonstrate the competition between $\alpha$-decay and SF. As an alternate method, machine learning algorithms are also used to probe the contest of $\alpha$-decay and SF for the range of 111$\leq$Z$\leq$118. These modified formulas i.e. modified Horoi formula (MHF-2020) (set-1 and set-2) and modified Sobiczewski formula (MSF-2020), along with machine learning methods i.e. MLP and XGBoost algorithm, provide an excellent match with available experimental data and half-lives. These formulas and algorithms are used to construct $\alpha$-decay chain of $^{286-302}$Og and $^{287-303}$119. Among them, $^{286-291}$Og and $^{287-296,302,303}$119 are found with a very long $\alpha$-decay chain (6$\alpha$/7$\alpha$) and reported here as potential candidates for future experiments in the superheavy domain.

\section*{ACKNOWLEDGMENTS}
Authors are thankful to Prof. H. F. Zhang and X. J. Bao for the valuable communication. Authors are indebted to Amit Garg, GWECA regarding the various discussion on machine learning. G. Saxena acknowledges the support provided by SERB (DST), Govt. of India under CRG/2019/001851.

\end{document}